# Local Polarization Switching in the Presence of Surface Charged Defects: Microscopic Mechanisms and Piezoresponse Force Spectroscopy Observations

Anna N. Morozovska[*], Sergei V. Svechnikov,

Institute of Semiconductor Physics, National Academy of Science of Ukraine,

45, pr. Nauki, 03028 Kiev, Ukraine

Eugene A. Eliseev

Institute for Problems of Materials Science, National Academy of Science of Ukraine,

3, Krjijanovskogo, 03142 Kiev, Ukraine

Brian J. Rodriguez, Stephen Jesse, and Sergei V. Kalinin[†]

[3]Materials Science and Technology Division and Center for Nanophase Materials Science,

Oak Ridge National Laboratory, Oak Ridge, TN 37831

**Abstract**

The thermodynamics and kinetics of tip-induced polarization switching in Piezoresponse Force Microscopy in the presence of surface charge defects is studied using the combination of analytical and numerical techniques. The signature of the defects in hysteresis loop fine structure and Switching Spectroscopy PFM images is identified and compared to experimental observations. An approach for the deconvolution of PFM spectroscopy measurements to extract relevant defect parameters is derived. This methodology is universal and can be extended to switching in other ferroics and in reversible electrochemical processes, establishing a pathway for the understanding of the thermodynamics and kinetics of phase transitions at a single defect level.

---

[*] Corresponding author, morozo@i.com.ua

[†] Corresponding author, sergei2@ornl.gov

# I. Introduction

Order parameter dynamics in ferroic materials, as well as processes such as electrochemical and solid state reactions, are strongly mediated by the presence of defects.[1,2,3] In ferroelectric materials, a bias-induced transition between two equivalent polarization states (180° switching) is reversible and is not associated with diffusion, mass transport, and significant heat exchange and strain effects. Combined with the atomic-scale width of the ferroelectric domain wall, this enables applications such as non-volatile random access memories,[4,5] ferroelectric tunnel junctions,[6,7] and high-density data storage.[8,9] These applications necessitate understanding of polarization switching in nanoscale volumes and elucidating the role of a single crystallographic or morphological defect on polarization switching, beyond the applicability limit of statistical theories.[10]

Similarly to other crystalline solids, ferroelectric and multiferroic crystals and films contain a range of point and extended defects in the bulk and at surfaces and interfaces. From simple energy considerations, the extended defects generally affect switching behavior stronger than localized ones. In polycrystalline materials and non-ideal single crystals, the switching is typically initiated at second phase inclusions, grain boundaries, microcracks, etc,[11] precluding unambiguous identification of defect types on the atomic level. Compared to polycrystalline materials, epitaxial thin films offer the advantage of better-understood defect structures, including threading and misfit dislocations,[12,13,14] the density of which can be tailored by a proper choice of deposition conditions and film thickness.[15] Recent advances in (Scanning) Transmission Electron Microscopy and atomically resolved Electron Energy Loss Spectroscopy[16] have brought the capability to probe the structure (e.g. direction and Burgers vector of dislocations) of a defect, and also determine the core atomic and electronic structures as well as dopant segregation,[17,18,19,20,21,22] thus determining the dislocation charge and dipole moment, i.e. the quantities that directly couple to ferroelectric polarization.

The role of defects on kinetics and thermodynamics of polarization switching, as well as other phase transitions, is threefold. Defects can determine local phase stability (e.g. shift the Curie temperature), act as nucleation centers in phase transitions, and pinning centers for moving transformation fronts. The defect contribution to properties can be analyzed on a statistically averaged level, i.e. the role of defect population on the effective thermodynamic and kinetic properties of the system.[1] Alternatively, changes in local materials properties induced by the electrostatic and elastic field of a defect, and its effect on local phase stability, domain wall pinning, and domain nucleation can be studied on a single-defect level.[2]

The role of defects on macroscopic phase stability has been studied extensively within the framework of Landau theory.[1,23] On a single-defect level, analysis requires the introduction of an appropriate structural model for the defect. The predominance of dislocations at the primary defect type in ferroelectric films has instigated a broad theoretical and experimental effort in determining their role on thermodynamic phase stability. The thermodynamic modeling by S. Alpay et al. and other groups (see e.g. Ref. 24) has demonstrated that dislocations locally destabilize ferroelectric phase,[25] and misfit[26] and threading[27] dislocation can thus account for ~10 nm non-switchable layer and reduced dielectric properties[28] in most ferroelectric films. This prediction was confirmed by the variable temperature electron microscopy studies by R. Wang et al.,[29] demonstrating the shift of ferroelectric transition temperature in the vicinity of dislocation. Specifically, dipole moment and charge of a dislocation directly favor one polarization orientation (random field), while strain fields can destabilize (or induce) ferroelectric phase and induce transition to the non-ferroelectric state (random bond), mapping the realistic defect structure on well-known statistical physics models.

The role of defects on domain wall pinning has been studied extensively both from statistical and local perspectives. From the statistical perspective, the role of random field and random bond defects on the domain wall dynamics and geometry in ferroelectric and ferromagnetic systems has been studied in detail.[30] Experimentally, the domain wall roughness was addressed by a series of papers by e.g. Tybell, Paruch et al.[31,32,33], and the pinning sites were attributed to oxygen vacancies. Misfit dislocations aligned in (100) and (010) directions effectively couple to the 90 degree domain walls (strain effects), and thus serve as effective pinning centers, as studied theoretically by Pertsev.[34] Atomic-scale studies by high-resolution electron microscopy allows domain wall positions to be correlated with structural defects, identifying the latter as pinning sites, as illustrated by Alexe et al.[35] for misfit dislocations. The effect of threading dislocations was addressed only at the macroscopic level.[36]

Finally, the role of defects on polarization switching in ferroelectrics has been recognized since the seminal work by Landauer 50 years ago,[37] stimulating half a century long effort to identify the defect types that affect switching and pinning. On the macroscopic level, the spatial and energy distribution of nucleation cites is a central element of Kolmogorov-Avrami[38,39] type theories of phase transitions.[40,41] A number of theoretical studies on domain nucleation on a local level[42,43,44,45] has been reported. Currently, atomistic studies of nucleation processes are being performed,[46] suggesting the atomistic mechanism of defect-mediated switching will become addressable. However, experimental studies of nucleation processes are significantly more challenging due to the low resolution of imaging techniques compared to nucleus size (~1-3 nm in ferroelectric) and the low concentration of nucleation centers. Recent studies by Grigoriev et al.[47] using ultrafast focused X-ray imaging, and Gruverman et al.[48] and Noh et al.[49] using piezoresponse force microscopy has demonstrated that in the uniform field created in ~100 micron capacitor structures, the switching is initiated in very few (~1-10) locations and then propagates through the macroscopic (~10s of microns) region of the film. While the process is reproducible and the defect locations can be determined repeatedly, their identity and the energetic parameters of the nucleation process are still an enigma. Furthermore, the rapid domain growth after nucleation in the uniform fields precludes observation of the early stages of the nucleation process.

This summary illustrates that despite the significant effort on studies of domain switching mechanisms in ferroelectrics, the key element required for linking macroscopic statistical theories and switching studies with atomically resolved microscopic imaging – the capability to probe the thermodynamics of the switching process on a single defect center – has been missing. Below, we discuss the applicability of spectroscopic imaging by Scanning Probe Microscopy (SPM) for visualizing defect centers and probe the thermodynamics and kinetics of local phase transitions on a single defect level.

## II. SPM studies of Local Phase Transitions

Scanning probe microscopy provides a natural framework for probing local phase transitions and correlating them with microstructure. In these measurements, the external stimulus (either local or global) applied to the systems induces phase transformation, while the SPM probe determines the change in local properties associated with the transition. Perhaps the best known example of such measurements is protein unfolding spectroscopy, in which the force applied by an AFM tip acts as a stimulus to change the molecule conformation, and the measured change in the molecule length provides readout.[50,51,52] Typically, the unfolding process is reversible, which allows determining the statistical distributions of the possible trajectories through the energy space of the system.[53] However, this example is unique – in cases such as pressure induced phase transitions (e.g. dislocation nucleation during indentation process) the process is irreversible, precluding the systematic studies of the role of defects on transition mechanisms.

To complement the force-induced phase transition, SPM allows bias-induced phase transitions to be studied. Unlike pressure, the probe bias can be made both positive and negative, allowing for the reversibility of the process. The ideal model system is the ferroelectric materials, in which polarization can be switched reversibly between the antiparallel states under the action of dc electric field. A significant insight into local switching processes in ferroelectrics has been achieved with the invention of Piezoresponse Force Microscopy. Here, the probe concentrates an electric field to a nanoscale volume of material (~10-50 nm) and induces local domain nucleation and growth. The size of the created domain as a function of length and duration of the switching pulse is imaged, providing information of switching process. Recent examples include studies by Ramesh and Waser,[54,55] Rosenman,[56] Kholkin,[57] and Hong,[58] demonstrating the scaling laws for bias-induced domain growth. These studies allow direct imaging of domain growth, but are (a) extremely time consuming (~ 10s hours/location, as compared to ~1s/spectrum), and (b) the smallest domains (corresponding to as-nucleated state can be below the resolution limit of the system. Complementary to these are the studies by Kholkin,[59] and Allegrini,[60,61] based on the statistical analysis of the domain patterns and domain wall roughness, which provide the information on the collective effect of defect centers on the switching process.

An alternative to the direct imaging are studies based on PFM spectroscopy. In these, the switching dc bias and probing ac bias are applied to the tip simultaneously. The probe detects the onset of nucleation and the size of a forming domain via detection of the electromechanical response. The resulting local electromechanical hysteresis loop contains information on local switching. In particular, loop fine structure (similar to fine structure on unfolding curves in force spectroscopy) is indicative of domain—defect interactions.[62]

Recently, Kalinin et al [63] demonstrated an approach to study the thermodynamics and kinetics of tip-induced nucleation processes using the fine structure analysis.

As compared to *structural* imaging by SPM, in which image morphology (pixel-to-pixel variation) provides information, *functional* SPM imaging and *spectroscopic mapping* brings a challenge of a quantitative data interpretation to interpret a value or spectrum acquired at a single point and its variations along the surface. This includes both semiquantitative analysis (e.g. determine the signatures of the defect in local spectroscopic data), and developing the quantitative relationship between the defect parameters and the measured signal. For studies of bias-induced local phase transitions in an SPM experiment, the following key elements can be delineated:

1. Determine the spatial distribution of the local driving force for the phase transition for known tip geometry

2. Analyze the energetic parameters of the phase transition in the non-uniform field and establish corresponding critical nucleus size in an ideal material

3. Determine the thermodynamics of the local transition in the presence of a defect

4. Establish the relationship between the size of phase-transformed region and the measured response for known tip geometry

5. Determine the tip geometry using appropriate calibration

Most of the individual steps in this scheme have already been demonstrated. Specifically, the exact solution of the PFM contact mechanics problem using an extension of the Hertzian contact problem has been demonstrated,[64,65] as were approximate solutions based on decoupling approximations (currently limited to point mechanical contact, corresponding to a weak indentation scenario when electrostatic fields generated outside the contact area dominate).[66,67,68,69] These yield the structure of electroelastic fields produced by the tip, (1). Similarly, an approach for tip calibration, (5), and the interpretation of spectroscopic data, (4), has been developed for special cases of tip geometry. The switching in an ideal material, (2), based on the point-charge model in the prolate ellipsoid geometry of Landauer[70] was pioneered by Molotskii *et al*.[71] This model was significantly extended by taking into account a finite tip size to determine the critical parameters of nucleation process by Morozovska et al.[72]

In this manuscript, we extend this analysis to develop a theoretical framework to describe the nucleation process in the vicinity of a surface field defect, relating the thermodynamics of a tip-induced phase transition to defect properties. We consider simple cases of well-separated surface field defects in semi-infinite material. This analysis lays the foundation for local studies of defect effects on phase transitions at a single defect level.

### III. General approach and problem statement
### III.1. General approach

Understanding the role of defects on polarization switching necessitates the analysis of the thermodynamics of the switching process. The free energy of a nucleating ferroelectric domain is

$$\Phi(\mathbf{r},U) = \Phi_{DS}(\mathbf{r},U) + \Phi_{Dom}(\mathbf{r},U) \qquad (1)$$

where the domain geometry is described by the *N*-dimensional parameter vector $\mathbf{r}$ and $U$ is electric bias applied to the local probe in the proximity with the surface.

The first term in Eq. (1) contains the contributions from frozen (defects, *d*) and thermal (*TD*) disorder, $\Phi_{DS}(\mathbf{r},t) = \Phi_d(\mathbf{r}) + \Phi_{TD}(\mathbf{r},t)$, within the volume of the domain. Note that disorder components can contribute differently to switching between different states, i.e. for 180° switching the energy $\Phi_d^+(\mathbf{r})$ for $+P \to -P$ is not necessarily equal to $\Phi_d^-(\mathbf{r})$ for $-P \to +P$. The symmetric and antisymmetric part of the frozen disorder components are referred to as effective random bond and random field disorder.

The second term in Eq. (1) is $\Phi_{Dom}(\mathbf{r},U) = \Phi_S(\mathbf{r}) + \Phi_p(\mathbf{r},U) + \Phi_D(\mathbf{r})$. It comprises the contributions of domain wall energy, $\Phi_S(\mathbf{r})$, the depolarization energy, $\Phi_D(\mathbf{r})$, and the interaction energy with probe tip electric field, $\Phi_p(\mathbf{r},U)$. The analysis of the switching process can be simplified for a rigid piezoelectric, for which effective materials constants are independent of the electric field. In this case, the interaction energy is $\Phi_p(\mathbf{r},U) = U\Phi_U(\mathbf{r})$. Note however that its equilibrium value is nonlinear with the tip bias $U$, since the corresponding domain parameters $\mathbf{r}$ (e.g. sizes) are voltage dependent and can be derived from Eq. (1).

The stochastic dynamics of the system described by Eq. (1) is well-studied in the context of chemical reactions[73] and protein unfolding spectroscopy.[50,51,52,53,74] Typical energy barriers for the polarization switching are much higher than thermal fluctuations in perfect ferroelectric materials (e.g. barrier is much greater than ~$10^3$ $k_BT$ for the plain electrode geometry). Hence, the thermal disorder and variability of switching behavior on repetitive switching cycles is anticipated to be negligibly small, and the equilibrium domain growth will proceed along the lowest free energy path. This is in agreement with high reproducibility of fine structure between the loops.[63]

The domain nucleation can be represented as a transition process on an *N*-dimensional surface of $\Phi(\mathbf{r},U)$. In the absence of defects, $\Phi(0,U)=0$. Due to the local nature of the electro-elastic field produced by the tip, $\Phi(\infty,U)=\infty$. Finally, the fact that electric field is finite in the vicinity of the tip-surface junction suggests that on the ideal surface, $\partial\Phi(\mathbf{r}=0,U)/\partial\mathbf{r}>0$. Alternatively, the domain nucleation will proceed spontaneously, corresponding to a different ground state of the system (surface state).

Stable domain configuration(s) correspond to local minima on the $\Phi(\mathbf{r},U)$ surface, where minima corresponding to $\Phi(\mathbf{r},U)>0$ are metastable and the ones with $\Phi(\mathbf{r},U)<0$ are stable. In the case of first order phase transitions, the minima and coordinate origin are separated by saddle point(s). The voltage *U* at which the stable minima (i.e. domain) appears is called the critical voltage, $U_{cr}$. The voltage of saddle point appearance, $U_{sp}$, corresponding to domain metastability, is usually close to $U_{cr}$.

The free energy value in the saddle point determines the activation energy, $E_a$, of domain nucleation. In the thermally induced nucleation limit, the domain nucleation process is analyzed as thermally activated motion in the phase space of the system along the minimum energy path connecting the origin and one of the local minima. The relaxation time necessary for the stable domain formation at $U_{cr}$ is maximal and the critical slowing down appears in accordance with general theory of phase transitions. Within the framework of activation rate theory, the domain nucleation takes place at higher activation voltage $U_a$ determined from the condition $\Phi(U_a)=E_a$, corresponding to the activation time $\tau=\tau_0 \exp(E_a/k_BT)$. For instance, the activation energy $E_a=20k_BT$ corresponds to a relatively fast nucleation time $\tau \sim 10^{-3}$ s for phonon relaxation time $\tau_0=10^{-12}$ s, while the condition $E_a \leq 2k_BT$ corresponds to "instant" or thermal nucleation.

The difference between the voltages corresponding to the formation of a saddle point, and a stable domain, $U_{sp} - U_{cr}$, determines the width of the (rather thin) thermodynamic hysteresis loop. More realistic models of piezoresponse hysteresis loop formation consider domain wall pinning effects. In the weak pinning limit, the domain growth in the forward direction is assumed to follow the thermodynamic energy minimum, while on decreasing bias, the domain remains stationary due to domain wall pinning by the lattice and atomic defects.

Experimental observations have demonstrated that nucleation voltages generally vary along the surfaces, indicating the presence of regions with reduced or increased nucleation potentials.[75] Furthermore, hysteresis loops often exhibit highly-reproducible fine structure.[63] This behavior can be attributed to defects below or at finite separation from the tip-surface junction. In the free energy space of the system, this suggests the presence of multiple minima separated by saddle points. Below, we analyze the polarization switching in the presence of field defects that couple to polarization.

### III.2. Problem statement

The first step in describing the defect-mediated phase transition is the construction of an effective defect potential that couples to the order parameter. Here we follow the approach of Gerra et. al.[45] assuming that the defect causes the built-in electric field, that directly couples to polarization (random field). Furthermore, we consider the nucleation process in a Landauer model for domain geometry[70] adapted by Molotskii[76] for tip-induced switching. The surface and electrostatic energy of the semi-ellipsoidal domain is:

$$\Phi(\mathbf{r},U) = \Phi_S(\mathbf{r}) + \Phi_D(\mathbf{r}) + \Phi_P(\mathbf{r},U) + \Phi_d(\mathbf{r}), \qquad (2a)$$

where the surface, depolarization, tip-induced, and defect contributions to free energy are

$$\Phi_S(\mathbf{r}) = \psi_S S \qquad (2b)$$

$$\Phi_D(\mathbf{r}) = \frac{n_D}{\varepsilon_0 \varepsilon_{11}} P_S^2 V \qquad (2c)$$

$$\Phi_P(\mathbf{r},U) = -2P_S \int_V d^3x \cdot E_3^p(\mathbf{x}) \qquad (2d)$$

$$\Phi_d(\mathbf{r}) = -2P_S \int_V d^3x \cdot E_3^d(\mathbf{x}) \qquad (2e)$$

and $S$ and $V$ are the domain surface and volume, $\psi_S$ is the domain wall energy density, $P_S$ is the magnitude of material spontaneous polarization $\mathbf{P} = (0, 0, P_S)$. The term Eq. (2c) is depolarization field energy calculated under the condition of perfect tip-surface electric contact or/and surface screening by free charges. The rigorous expression for the depolarization factor, $n_D$, is given in Ref. [77] for an ellipsoidal domain shape. The electric field established by the probe is $\mathbf{E}^p(\mathbf{x}) = -\nabla \varphi_p(\mathbf{x})$, and the electric field created by the defects is $\mathbf{E}^d(\mathbf{x}) = -\nabla \varphi_d(\mathbf{x})$.

Further analysis is performed assuming that the semi-ellipsoidal domain is axi-symmetric, i.e. it has radius $r$ and length $l$, but allowing for defect influence the domain center is shifted on value $y_0$ compared to the tip location. The center of the nearest surface field defect is assumed to be located at position $\mathbf{x}_0 = \{x_{10}, 0, 0\}$, whereas the tip is located at the coordinate origin (see Fig.1).

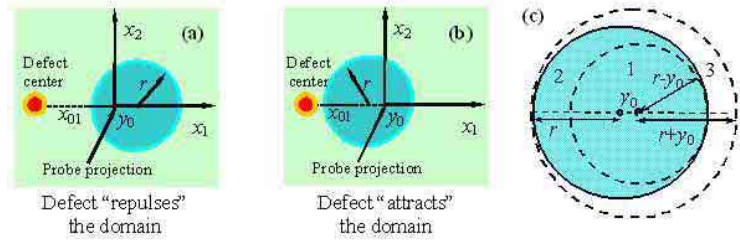

**FIG. 1**. Domain nucleation in the vicinity of surface defect with center located in the point $\{x_{01}, 0, 0\}$. (c) Scheme of the defect-induced PFM response change estimation, the ratio of the area 2 to the area 3 is $(2r - |y_0|)/(2r + |y_0|)$.

Rigorously speaking, the domain shape will deviate from semi-ellipsoidal near the defect (the system radial symmetry is broken). However, analytical treatment of the problem and necessity to calculate the depolarization field exactly imposes a limitation on the number of free parameters describing geometry. Hereinafter, we neglect the "shape asymmetry", but consider the domain center displacement as a variational parameter.

### III.3. Domain-probe interaction energy

For transversally isotropic material and a rotationally symmetric probe, corresponding to the prototype case of switching in tetragonal and hexagonal ferroelectrics, the probe electric potential $\varphi_p(\mathbf{x})$ is an axially symmetric even function, i.e. $\varphi_p(\mathbf{x}) \equiv \varphi_p(\sqrt{x_1^2 + x_2^2}, x_3)$. In this case, the domain-tip interaction energy $\Phi_p(r, l, y_0)$ can be rewritten as

$$\Phi_p(r, l, y_0) = \int_S 2\mathbf{P}_S \varphi_p(x_1 + y_0, x_2, x_3) d\mathbf{s} = \int_S 2\mathbf{P}_S \sum_{m=0}^{\infty} \frac{\partial^{2m} \varphi_p(x_1, x_2, x_3)}{\partial x_1^{2m}} \frac{y_0^{2m}}{(2m)!} d\mathbf{s} \qquad (3)$$

It is clear that the symmetry $x_1 \leftrightarrow x_2$ exists in Eq. (3). Using Gauss theorem we obtain from series Eq. (3) that:

$$\Phi_p(r, l, y_0) \approx 4\pi P_S \int_0^l dx_3 \int_0^{r\sqrt{1-x_3^2/l^2}} \rho d\rho \left( E_3^p(\rho, x_3) + \frac{y_0^2}{4\rho} \frac{\partial}{\partial \rho}\left(\rho \frac{\partial}{\partial \rho} E_3^p(\rho, x_3)\right) + ...\right). \qquad (4)$$

Here $\sqrt{x_1^2 + x_2^2} = \rho$ and $x_3$ is the radial and vertical coordinate respectively; $E_3^p(\rho, x_3) = -\partial \varphi_p(\rho, x_3)/\partial x_3$ is the longitudinal field component.

The flattened or spherical probe potential $\varphi_p$ can be modeled using an effective point charge approximation. The probe is represented by a single charge $Q$ located at distance $d$ from a sample surface (see details in Ref. [78]). The potential $\varphi_p$ at $x_3 \geq 0$ has the form:

$$\varphi_p(\rho, x_3) \approx \frac{U d}{\sqrt{\rho^2 + (x_3/\gamma + d)^2}}. \qquad (5)$$

Here $U$ is the bias applied to the probe, $\gamma = \sqrt{\varepsilon_{33}/\varepsilon_{11}}$ is the dielectric anisotropy factor. In the case of local point charge model, the probe is represented by a single charge $Q = 2\pi\varepsilon_0\varepsilon_e R_0 U (\kappa + \varepsilon_e)/\kappa$ located at $d = \varepsilon_e R_0/\kappa$ for a spherical tip, or $d = 2R_0/\pi$ for a flattened tip represented by a disk in contact. Here, $\kappa = \sqrt{\varepsilon_{33}\varepsilon_{11}}$ is the effective dielectric constant.

Substituting potential Eq. (5) into the series Eq. (4) and performing the integration, we derive the Pade approximation for the tip-induced interaction energy for a shifted domain as:

$$\Phi_p(r,l,y_0) \approx \frac{-4\pi U P_S\, d\, r^2\, l/\gamma}{\left(\sqrt{r^2+d^2+y_0^2}+d\right)\left(\sqrt{r^2+d^2+y_0^2}+d+l/\gamma\right)}. \tag{6}$$

### III.4. Domain-defect interaction energy

Assuming that the defect-induced built-in electric field can be represented as $E_3^d(\mathbf{x}) = E_3^d\!\left((x_1-x_{01})^2 + x_2^2, x_3\right)$, i.e. it is an axi-symmetric even function with respect to the defect center, the domain-defect interaction energy is

$$\Phi_d(r,l,y_0) = -2P_S \int_V d^3x\, E_3^d(x_1 - x_{01} + y_0, x_2, x_3), \tag{7}$$

where the integration is performed within the volume $(x_1^2+x_2^2)/r^2 + x_3^2/l^2 \leq 1$. The defect contribution to the free energy of a domain can be rewritten via the overlap integral:

$$\Phi_d(r,l,x_{01}-y_0) = -2P_S \int_0^l dx_3 \int_0^{2\pi} d\varphi \int_0^{r\sqrt{1-x_3^2/l^2}} \rho d\rho\, E_3^d\!\left(\rho^2 + (x_{01}-y_0)^2 - 2\rho\cos\varphi(x_{01}-y_0),\, x_3\right) \tag{8}$$

Further analysis depends on the defect model, i.e. the distribution of the built-in electric field, $E_3^d$. In Section IV we calculate the energy Eq. (8) for several types of surface defects.

### III.5. Effective piezoresponse calculations

Measured in a PFS experiment is the electromechanical response related to the size of ferroelectric domain formed below the tip. Hence, to calculate the shape of the PFM hysteresis loop, the electromechanical response change induced by the semi-ellipsoidal domain is required. Within the framework of linearized theory by Felten et al.[66] the surface displacement vector $u_i(\mathbf{x})$ at position $\mathbf{x}$ is

$$u_i(\mathbf{x}) = \int_0^\infty d\xi_3 \int_{-\infty}^\infty d\xi_2 \int_{-\infty}^\infty d\xi_1\, \frac{\partial G_{ij}(\mathbf{x},\xi)}{\partial \xi_l} E_k^p(\xi) d_{kmn}(\xi,r,l,y_0) c_{nmjl} \tag{9}$$

where $\xi$ is the coordinate system related to the material, $d_{nmp}$ are strain piezoelectric coefficients distribution, $c_{nmjl}$ are elastic stiffness and the Einstein summation convention is used. $E_k^p$ is the electric field created by the biased probe, derived from Eq. (5). Note that the electric field distribution that induces domain switching and that determined the detection mechanisms are the same. Hence, the problem is strongly non-linear in electric field, necessitating the mathematical analysis developed below. For typical ferroelectric perovskites, the symmetry of the elastic properties can be approximated as cubic (anisotropy of elastic properties is much smaller then that of dielectric and piezoelectric properties) and therefore an isotropic approximation can be used for the Green's function $G_{ij}(\mathbf{x}, \xi)$.[68, 79]

Integration of Eq. (9) for $x_3 = 0, \rho = 0$ yields the expression for effective vertical piezoresponse, $d_{33}^{eff} = u_3/U$, as

$$d_{33}^{eff}(r,l,y_0) = d_{31} g_1(r,l,y_0) + d_{15} g_2(r,l,y_0) + d_{33} g_3(r,l,y_0), \tag{10}$$

where $g_i = f_i - 2w_i$, functions $w_i = 0$ in the initial and $w_i = f_i$ in the final state. The functions $f_i$ are $f_1 = (2(1+\gamma)\nu + 1)/(1+\gamma)^2$, $f_2 = -\gamma^2/(1+\gamma)^2$, $f_3 = -(1+2\gamma)/(1+\gamma)^2$ and define the electromechanical response in the initial and final states of switching process.[80] In this approximation, the relevant materials properties are the Poisson ratio, $\nu$.

The functions $w_i$ are dependent on the domain sizes $r$, $l$ and domain shift with respect to the tip apex, $y_0$. Considering the signal generation volume in PFM, we argue that piezoresponse changes negligibly, $|w_i| << |f_i|$, when the domain is far from the tip, e.g. under the condition $|y_0| >> r$. For the opposite case $|y_0| < r$, functions $w_i$ have relatively simple integral representations:

$$w_1(r,l,y_0) = \frac{1}{2\pi} \int_0^{2\pi} d\varphi \int_0^{\pi/2} d\theta (3\cos^2\theta - 2(1+\nu))\cos\theta \sin\theta \frac{R_w(\theta,\varphi,r,l,y_0)}{R_G(\theta,\varphi,r,l,y_0)}, \tag{11a}$$

$$w_2(r,l,y_0) = \frac{3}{2\pi} \int_0^{2\pi} d\varphi \int_0^{\pi/2} d\theta \left( \frac{\gamma d + \cos\theta R_w(\theta,\varphi,r,l,y_0)}{R_G(\theta,\varphi,r,l,y_0)} - 1 \right) \cos^2\theta \cdot \sin\theta, \tag{11b}$$

$$w_3(r,l,y_0) = -\frac{3}{2\pi} \int_0^{2\pi} d\varphi \int_0^{\pi/2} d\theta \cos^3\theta \sin\theta \frac{R_w(\theta,\varphi,r,l,y_0)}{R_G(\theta,\varphi,r,l,y_0)}. \tag{11c}$$

Here, the radius $R_w(\theta,\varphi,r,l,y_0)$ determines the domain wall shape and its center position. In the typical case of prolate semiellipsoid ($r << l$) or cylinder we derive

$$R_w(\theta,\varphi,r,y_0) = \frac{|y_0|\cos\varphi + \sqrt{r^2 - y_0^2 \sin^2\varphi}}{\sin\theta}. \tag{12a}$$

The function $1/R_G(\theta,\varphi,r,y_0)$ is related to the probe electrostatic potential in the domain wall point determined by $R_w(\theta,\varphi,r,y_0)$, namely

$$R_G(\theta,\varphi,r,y_0) = \sqrt{(\gamma d + \cos\theta\, R_w(\theta,\varphi,r,y_0))^2 + \gamma^2 \sin^2\theta\, R_w^2(\theta,\varphi,r,y_0)}. \quad (12b)$$

At $y_0 = 0$ (no lateral shift) expressions Eq. (11a-c) coincide with the ones derived for domain nucleating on the tip axis in Ref. [72], as expected.

Using approximate expressions derived for $y_0 = 0$ in Ref. [69], approximate analytical relationship between the radius of a prolate semiellipsoidal domain, $r$, lateral shift, $|y_0|$, and the PFM signal can be determined as the superposition of the inner cylindrical domain with radius $(r - |y_0|)$ and the part of the ring with inner radius $r_i = (r - |y_0|)$ and outer radius $r_o = (r + |y_0|)$ (see Fig.1(c)):

$$d_{33}^{eff}(r, y_0) = d_{33}^{dom}(r - |y_0|) + d_{33}^{ring}(r - |y_0|, r + |y_0|)\frac{2r - |y_0|}{2r - |y_0|}, \quad (13a)$$

$$d_{33}^{dom}(r) = \frac{3}{4}d_{33}^* \frac{\pi d - 8r}{\pi d + 8r} + \frac{d_{15}}{4}\frac{3\pi d - 8r}{3\pi d + 8r}, \quad (13b)$$

$$d_{33}^{ring}(r_i, r_o) = d_{33}^{dom}(r_o) - d_{33}^{dom}(r_i). \quad (13c)$$

Here, the material is approximated as dielectrically isotropic, $\gamma \approx 1$, $r = r(U)$ is the voltage dependent domain radius, and $d_{33}^* = d_{33} + (1 + 4\nu)d_{31}/3$.

## IV. Surface field defects
### IV.1. Domain free energy affected by a surface field defect

On the structural level, defects in ferroelectric materials are associated with the disruption in lattice periodicity and associated changes in electronic structure. Local charge redistribution in the defect core is compensated by local bend bending and Debye screening, leading to the exponential vanishing of Coulomb electric fields away from the localized defect. Far from structurally-distorted defect core, the long-range electric field couples linearly to the polarization order parameter, stabilizing preferential polarization states. Therefore, the choice of electric field, rather than charge distribution, as a starting model for the defect is motivated by (a) the fact that field, rather then charge, couples to the polarization,

(b) the field-distribution models are more universal and less sensitive to the exact atomic and electronic structure of the defect, and (c) short-form analytical expressions can be obtained. While the numerical analysis can be performed for arbitrary charge distributions (from which field structure can be reconstructed), resulting complex expressions are not amenable to analytical treatment.

*IV.1.1. Dislocation-surface junction (Type I) and surface dipole patch (Type II) defects*

Here we adopt a model of a well-localized surface field defect with characteristic radius, $r_d$, and penetration depth $h_d \ll r_d$, located at the point $\{x_{01},0,0\}$ (see Fig. 2a). Based on comparison with the critical nucleation domain size and switching fields, the relevant values of the penetration depth are $h_d \sim 1-2$ nm, maximal field strength $E_S = 10^8 - 10^{10}$ V/m [Ref. 45] and defect radius is $r_d \sim 1-50$ nm. For larger radius, the defect becomes significantly larger than PFM tip size, and hence can be approximated by the homogeneous surface field considered in Ref. [45], while for a smaller defect size, effects on nucleation are minimal.

To develop the analytical description of defect-mediated switching, we consider a laterally localized defect, in which longitudinal component of electric field is

$$E_3^d(\mathbf{x}) = E_S f(x_3) \exp\left(-\frac{(x-x_{01})^2 + x_2^2}{r_d^2} - \frac{x_3}{h_d}\right) \quad (14)$$

Below we consider two limiting cases, $f(x_3) = 1$, and $f(x_3) = 1 - x_3/h_d$. These lead to qualitatively different behavior under the condition $|x_{01} - r_d| \lesssim d$ corresponding to the noticeable interference of charged probe and defect electric fields. Note that in the other limit, $|x_{01} - r_d| \gg d$, the tip is well-separated form the defect and hence the role of tip-induced switching is minimal. Therefore, here we analyze the switching behavior for $|x_{01} - r_d| \lesssim d$.

To establish the relationship between the field structure and the corresponding physical model, we reconstruct the charge density distributions corresponding to Eq. (14) for linear dielectric case below. The defect charge density $\sigma_d(\mathbf{x})$ can be found from the Maxwell equation $\varepsilon_0 \text{div}(\hat{\varepsilon} \mathbf{E}^d) = \sigma_d$ supplemented with the boundary condition $E_{1,2}^d(x_3 = 0) = 0$, corresponding to the full screening of the electric field on the sample surface (see Appendix A for details). The charge distribution $\sigma_d(\mathbf{x})$ related to the field distribution Eq. (14) is shown for $f(x_3)=1$ in Fig. 2(c-e) and for $f(x_3)=1-x_3/h_d$ in Fig. 2(f-g). It is clear that both distributions are maximal near the surface and rapidly decrease with the depth, $x_3$.

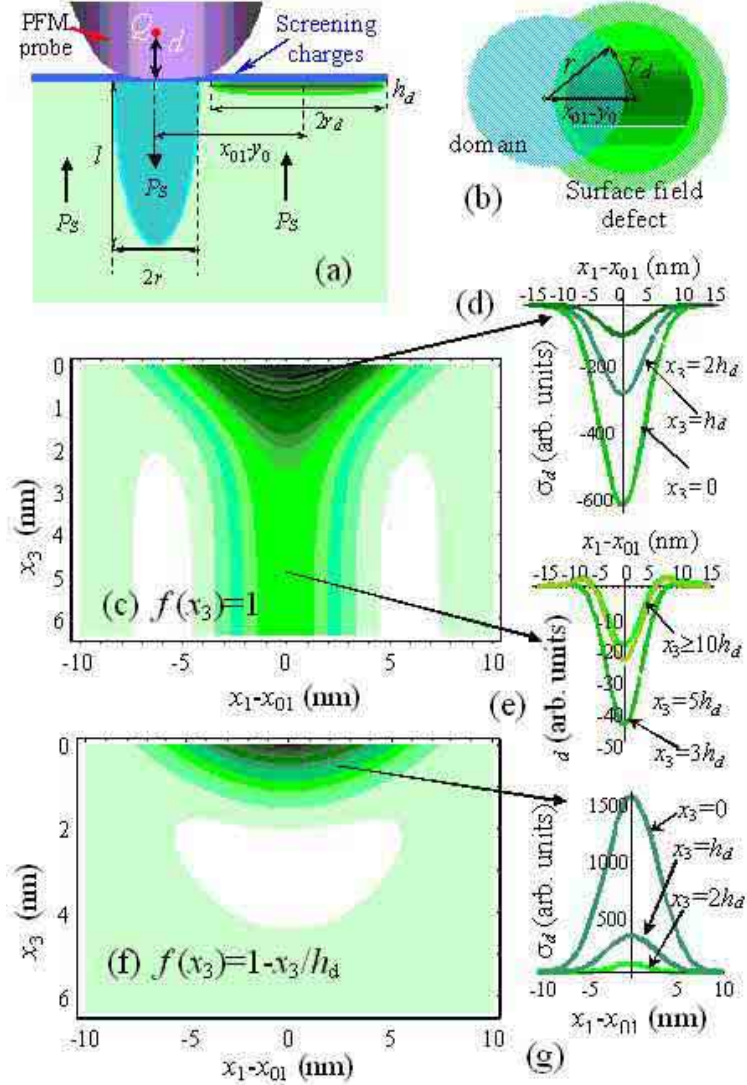

**FIG. 2**. Domain nucleation in the vicinity of surface field defect with characteristic radius $r_d$ located in the point $\{x_{01},0\}$. (a) Side view, (b) – top view. (c,f) Contour map of the defect charge density $\sigma_d(\mathbf{x})$ created by the field defect with radius $r_d = 4$ nm and penetration depth $h_d = 0.8$ nm. (d, e, g) Corresponding defect charge density $\sigma_d(\mathbf{x})$ cross-sections.

For the case $f(x_3)=1$, the density $\sigma_d(\mathbf{x})$ trends to a constant value at $x_3 \to \infty$ as shown in Fig. 2 (e). Such $\sigma_d(\mathbf{x})$ is a continuous charge distribution including surface spot and vertical charged line, while corresponding built-in field is a well-localized spot. The distribution could be related with charge accommodation at the *vertical dislocation line* and *dislocation-surface junction*, in agreement with expected behavior for threading dislocations in polar materials.[81]

The case $f(x_3)=1-x_3/h_d$ corresponds to the well-localized charge spot with exponentially vanishing charge density as shown in Fig. 2(f-g), and can approximate the case of a *surface dipole patch*, e.g. due to contaminations.

Substituting defect-induced electric field from Eq. (14) into the domain free energy given by Eq. (2)-(8), we obtain:

$$\Phi_S(r,l) = \pi \psi_S \, l \, r \left( \frac{r}{l} + \frac{\arcsin\sqrt{1-r^2/l^2}}{\sqrt{1-r^2/l^2}} \right) \tag{15a}$$

$$\Phi_D(\mathbf{r}) = \frac{P_S^2}{\varepsilon_0 \varepsilon_{33}} \frac{4\pi r^2 l}{3} \frac{(r\gamma/l)^2}{1-(r\gamma/l)^2} \left( \frac{\operatorname{arctanh}\left(\sqrt{1-(r\gamma/l)^2}\right)}{\sqrt{1-(r\gamma/l)^2}} - 1 \right) \tag{15b}$$

$$\Phi_P(r,l,y_0,U) = -\frac{4\pi U \, P_S \, d \, r^2 \, l/\gamma}{\left(\sqrt{r^2+d^2+y_0^2}+d\right)\left(\sqrt{r^2+d^2+y_0^2}+d+l/\gamma\right)} \tag{15c}$$

$$\Phi_d(r,l,x_{01},y_0) = -2\pi r_d^2 h_d P_S \, E_S I_S(r,l,x_{01}-y_0) \tag{15d}$$

The dimensionless overlap integral $I_S(r,l,x)$ has the form

$$I_S(r,l,x) = 2 \int_0^{l/h_d} dz f(z) \exp(-z) \int_0^{r/r_d} \rho d\rho \cdot I_0\left(2\frac{x}{r_d}\rho\right) \exp\left(-\rho^2 - \frac{x^2}{r_d^2}\right) \approx$$

$$\approx \begin{cases} \left(1-\exp\left(-\frac{l}{h_d}\right)\right)\left(1-\exp\left(-\frac{r^2}{r_d^2}\right)\right)\exp\left(-\frac{x^2}{r_d^2}\right), & f(z)=1 \\ \\ \frac{l}{h_d}\exp\left(-\frac{l}{h_d}\right)\left(1-\exp\left(-\frac{r^2}{r_d^2}\right)\right)\exp\left(-\frac{x^2}{r_d^2}\right), & f(z)=1-z \end{cases} \tag{16}$$

The scaling of the overlap integral Eq. (16) with domain length results in a qualitative difference between the defect effect on switching in the *charged dislocation line* and *dipole patch* cases. The defect contribution to domain energy, $I_S(r,l,x)$, is maximal at $l \to \infty$ for the *dislocation line* ($f = 1$), while it is maximal at $l \cong h_d$ and exponentially vanishing at $l \to \infty$ for the *dipole patch* ($f(x_3) = 1 - x_3/h_d$). The origins of this behavior are obvious from Figs. 2 (c-g), taking into account that the domain depolarization field vanishes as $1/l^2$ at $l \to \infty$. The interaction energy between the vanishing depolarization field with a well-localized patch tends to zero at $l \to \infty$, while the interaction energy between the vanishing depolarization field with the vertical charge line tends to a constant value. At $l \ll h_d$, the overlap integral coincides for both $f = 1$ and $f = 1 - x_3/h_d$. Thus, the overlap contribution to the free energy is qualitatively different for the considered cases: for the case $f = 1$ it could be essential at all values $l > h_d$, while for the case $f(x_3) = 1 - x_3/h_d$ noticeably smaller nonzero values are possible only within the range $0.2 h_d \leq l \leq 2 h_d$ that typically corresponds to an ultra-short domain, since $h_d \sim 1 - 2$ nm. For longer domains the defect influence on their formation is negligible for the case of the *surface dipole patch*.

After minimization of Eqs. (15) with respect to the domain center shift towards the defect, $y_0$, the domain free energy can be represented as a 2D surface in coordinates $r$ and $l$ (which then corresponds to the section of the full 3D surface). We obtained that $y_0$ is voltage and size dependent, i.e. $y_0 \equiv y_0(U, r, l)$. In particular, the shift $y_0$ has different values for nucleus and critical domain size, i.e. $y_0(U, r_S, l_S) \neq y_0(U, r_{cr}, l_{cr})$. The nucleus size $\{r_S, l_S\}$, minimal critical domain size $\{r_{cr}, l_{cr}\}$ and equilibrium size voltage dependences $\{r(U), l(U)\}$ are found from the free energy saddle point and minima, correspondingly.

To illustrate this behavior, below we compare the free energy contour maps, activation barrier, critical voltage and domain size that correspond to the states with different $x_{01}$ (i.e. tip positioned at different separations from the defect).

*IV.1.2. Defect center below the probe apex*

Even in the simplest case when the surface field defect is located just below the tip apex (i.e. $x_{01}=0$), the driving electric field spatial distribution appeared rather complex depending on the surface field amplitude $E_S$, its sign and halfwidth, $r_d$. On the sample surface, the total electric field can be written as

$$E_3(\rho,0) = \frac{U d^2}{\gamma(\rho^2+d^2)^{3/2}} + E_S \exp\left(-\frac{\rho^2}{r_d^2}\right), \qquad (17a)$$

as illustrated in Fig.3.

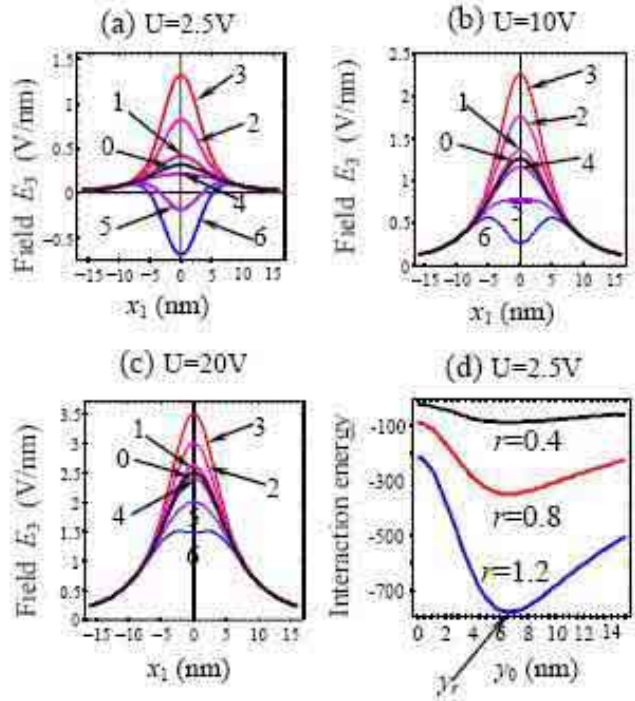

**FIG. 3**. Driving electric field $E_3(\rho,0)$ distribution on the surface at different applied voltages $U$: 2.5V (a), 10V (b), 20V (c). Surface field defect of radius $r_d = 4$ nm, penetration depth $h_d = 0.8$ nm, amplitude $E_S = 0; 10^8; 5\cdot 10^8; 10^9; -10^8; -5\cdot 10^8; -10^9$ V/m (curves 0-6) is located below the tip apex ($x_{01}=0$). (d) Corresponding interaction energy via the shift $y_0$ for $U=2.5$V, $E_S=-10^9$ V/m, $l=10$ nm and different domain radius $r=0.4; 0.8; 1.2$ nm (figures near the curves). Material parameters $P_S = 0.5$ C/m$^2$ and $\gamma \approx 1$, point charge-surface separation $d=8$ nm.

A number of interesting behaviors can be predicted depending on the relative magnitude of tip-induced and defect-induced electric fields as a result of the different distance behavior of the field components. For a positive surface field ($E_S > 0$) and a positive tip bias ($U > 0$) the field $E_3(\rho,0)$ is maximal just below the tip ($\rho = 0$) and so the domain forms exactly at $y_0 = x_{01} = 0$ (see curves 1-3 in Figs.3a-c). The situation is quantitatively the same for small enough negative surface field amplitudes $-U/\gamma d < E_S \leq 0$ (see curves 0, 4 in Figs.3a-c).

For positive bias and negative surface defect of sufficient field strength, $E_S < -U/\gamma d$, the driving electric field $E_3(\rho,0)$ could be maximal on the ring $\rho = y_r \neq 0$ at $x_3 = 0$ (see Fig.3a-b, curves 5, 6). Thus every point of the ring is an equal-probable candidate for domain nucleation. For instance, the domain-tip-defect negative interaction energy $\Phi_d(r,l,y_0) + \Phi_p(r,l,y_0)$ depicted in Fig. 3 (d) is minimal at $y_0 \approx 6$ nm and maximal at $y_0 \approx 0$ for nucleus radius and typical material parameters similar to those above. The ring-like domain nucleus cannot be treated quantitatively, since the chosen trial shape is semi-ellipsoid with circular cross-section.[82]. However, for large enough biases (2-3V for chosen material parameters) the interaction energy becomes minimal at $y_0 \approx 0$ with domain radius increases up to 10-20 nm, indicating that the center of the stable semi-ellipsoidal domain with a radius of more than several $r_d$ should be located below the tip. Thus, the lowest thermodynamic path of domain formation effected by the strong negative surface field defect located at $x_{01} = 0$ is expected to start on the ring $\rho = y_r \neq 0$ (nucleation stage) and then transforming into the stable domain with center at $y_0 \approx 0$. In Appendix B we obtained that at voltages $U < -2\sqrt{2}\gamma E_S d^3 / r_d^2$ the ring radius is $y_r \approx r_d \sqrt{\ln(-2\sqrt{2}\gamma E_S d^3 / U r_d^2)}$ for nucleus length $l_S \leq h_d \leq \gamma d$ and $y_r \approx r_d \sqrt{\ln(-\sqrt{2}\gamma E_S d^2 h_d / U r_d^2)}$ for $l \gg h_d$ and $l \gg \gamma d$. We expect that the ring-like domain may be stable for strong negative surface fields in the absence of fluctuations, i.e. when the corresponding activation voltage is essentially less than the value $-2\sqrt{2}\gamma E_S d^2 h_d / r_d^2$.

Voltage–dependent free energy surfaces defined by Eq. (15) in the presence of a surface field defect located directly below the tip apex ($x_{01}=0$) are shown in Figs. 4,5. The maps are calculated for PZT-6B ceramics (modified Pb(Zr,Ti)O$_3$ solid solution) and surface field defects created by a dislocation line ($f=1$) and a dipole patch ($f(x_3)=1-x_3/h_d$), correspondingly.

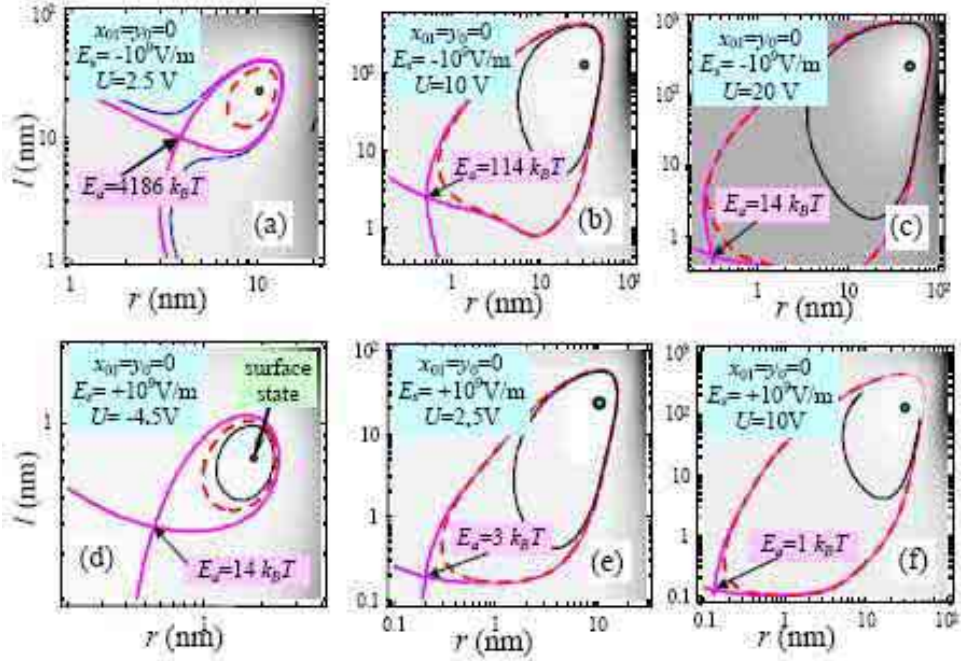

**FIG. 4**. Voltage dependence of free energy surfaces in the presence of surface field defect located below the tip apex ($x_{01}=0$) with $f=1$, radius $r_d=4$ nm, penetration depth $h_d=0.8$ nm, amplitude $E_S=-10^9$ V/m (parts a-c) and $E_S=+10^9$ V/m (parts d-f). Labels correspond to the domain shift $y_0$ in nm and activation energy values in $k_BT$ units. Dashed contour corresponds to zero energy. Small circle and arrow with label denote absolute minimum (equilibrium domain sizes) and activation barrier $E_a$ (saddle point and contour) correspondingly. Material parameters correspond to PZT-6B: $P_S=0.5$ C/m$^2$, $\varepsilon_{33}\approx 500$, $\gamma\approx 1$, $\psi_S=150$ mJ/m$^2$; point charge-surface separation $d=8$ nm corresponds to the local charge approximation for sphere-plane model of tip-surface contact.

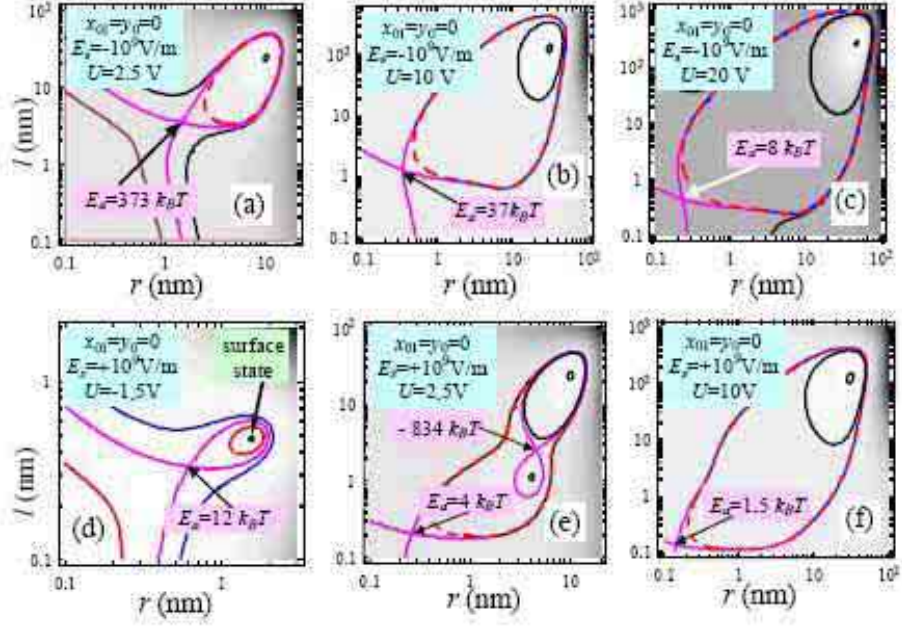

**FIG. 5**. Voltage dependence of free energy surfaces in the presence of surface field defect located below the tip apex ($x_{01} = 0$) with $f(x_3) = 1 - x_3/h_d$, $r_d = 4$ nm, $h_d = 0.8$ nm, positive maximal field $E_S = -10^9$ V/m (parts a-c) and negative field $E_S = +10^9$ V/m (parts d-f). Material parameters, tip characteristics and designations are the same as in Fig. 4.

Shown in Figs. 4,5 are the cross-section of $\Phi(r, l, y_0)$ in coordinates of domain radius, $r$, and length, $l$, at $y_0 = 0$. As shown above, the case $y_0 = 0$ is the nucleation site for voltages $U > -\gamma\, dE_S$ that corresponds to $U > 8$ V for $E_S = -10^9$ V/m and $U > -8$ V at $E_S = +10^9$ V/m for chosen materials and tip parameters. Thus the saddle points in parts (b)-(f) correspond to the lowest activation barrier, whereas the saddle shown in part (a) for $U = 2.5$ V corresponds to the highest barrier. Shown in Fig. 4a is the saddle at $y_0 = 0$ corresponding to the barrier of 4200 $k_BT$. The lowest saddle (with activation energy 2600 $k_BT$) appeared at $y_0 \approx 8$ nm (not shown, since it corresponds to the ring-like nucleus).

Similar to switching on a defect-free surface,[72] the activation barrier rapidly decreases with applied voltage. A favorable (positive) surface field defect decreases the activation barrier and thus stimulates domain nucleation at lower applied voltages [Figs. 4 (d-f) and 5(d-f)] in comparison with an unfavorable (negative) field defect [Figs.4 (a-c) and 5(a-c)].

The numerical estimates indicate that activation voltages (corresponding to the case when thermal nucleation is possible) are typically much greater than the critical ones (corresponding to the thermodynamic stability of the domains). For example, for the chosen material parameters and a negative surface field ($E_S < 0$), the critical voltage $U_{cr}^-$ determined from the condition $\Phi(U_{cr}^-) = 0$ at minimum, is about 2V, whereas the activation voltage $U_a^-$, determined from the condition $\Phi(U_a^-) \approx 20\,k_B T$ in a saddle point, is about 20V (compare parts (a) and (c) of Figs.4-5). In other words, the domain becomes thermodynamically stable at 2V, while the activation barrier $E_a$ becomes low enough for the process to be thermally-activated in a reasonable time only at 20V.

A positive field defect ($E_S > 0$) acts as a nucleation center at zero or even negative voltages (see e.g. Figs.4,5 d-e). The surface state (i.e. stable domain with length $l \cong h_d$, radius $r \leq r_d$ and center at $y_0 = x_{01}$) appears when the defect field strength exceeds the critical value, $E_S^{cr}$, determined as

$$E_S^{cr} \approx \frac{2e}{3(e-1)}\left(\frac{\psi_S}{h_d P_S} + \frac{P_S}{\varepsilon_0 \varepsilon_{11}}\right) \qquad (17b)$$

for a charged dislocation line ($f = 1$) or

$$E_S^{cr} \approx \frac{2}{3e}\left(\frac{\psi_S}{h_d P_S} + \frac{P_S}{\varepsilon_0 \varepsilon_{11}}\right) \qquad (17c)$$

for a dipole surface patch ($f = 1 - x_3/h_d$) (see Appendix B for details). The surface state becomes unstable for a negative external field $E_0 < -E_S^{cr}$. Note that in general, the surface state origin is similar to the domain nucleation in the tip-field. Experimentally, the surface state will correspond to a "frozen" polarization level for low enough fields.

Depending on the material parameters and surface field model, the surface state either extends or shifts up to the tip under the voltage increase, or two minimums appear as shown Fig. 5 (e). Numerical calculations have shown that the bistability (multiple minima) is more pronounced for $f = 1 - x_3/h_d$ and/or a small surface energy, $\psi_S$. For PZT-6B with a surface field characterized by $f = 1$, the surface state disappears only at $U_S^+ < -5$ V; for $f = 1 - x_3/h_d$ it happens at $U_S^+ < -1.5$ V. Using the activation level of $20 k_B T$, we obtained that the difference between the surface state voltage $U_S^+ \approx -(1-5)$ V for the positive field defect and the nucleation voltage $U_a^- \approx +20$ V for the negative one (at $3k_BT$) is more than 20V.

*IV.1.3. Defect at the intermediate separation from the tip apex*

The presence of the off-center defects gives rise to a rich spectrum of phase-transition behaviors depending on defect strengths, sign, and defect-tip separation. The effect of domain attraction or repulsion by the surface field defect with the center located at different distances $x_{01} \neq 0$ from the probe apex is illustrated in Fig. 6. Numerical calculations show that positive field defects with sufficient electric field strength, $E_S > E_S^{cr}$, located at distances $|x_{01}| \leq r_d$ always act as nucleation centers at voltages $U \geq 0$. During this process, $y_0 \approx x_{01}$ in the saddle point and the positive difference $(x_{01} - y_0)$ slightly increases with applied voltage increase (see insets indicating defect-induced nucleation). Even at high voltage the nucleus position is centered at the defect. However, the equilibrium domain position is below the tip, i.e. $y_0 \approx 0$ under the same other conditions (see main plots indicating tip-induced growth).

This analysis implies that the domain nucleus originated below the defect ($y_0 \approx x_{01}$ in the saddle point) and rapidly moves towards its equilibrium location below the tip ($y_0 \approx 0$) when the probe electric field substantially overcomes the defect field. Under certain conditions, the multiple minima corresponding to the domain position below the defect and below the tip appear for the case of $f = 1 - x_3/h_d$, as shown in Fig. 6 (d).

Negative defects with sufficient field strength, $E_S < -10^8$ V/m, located at distances $|x_{01}| \leq r_d$ always delay the nucleation [Figs. 6(c,e)], i.e. the domain nucleus repulses from the defect, $y_0 \leq 0$ (see corresponding saddle points). The repulsion is slightly stronger for $f = 1$ than for $f = 1 - x_3/h_d$ (compare $y_0$ values in the saddle points of (c) and (e)). Similarly to the case of positive field defect, the equilibrium domain position is below the tip, i.e. $y_0 \approx 0$ under the same other conditions (see main plots). This means that the domain nucleus repulsed by the defect originates far from the probe apex. Then the domain rapidly grows in the probe field towards its equilibrium location below the tip.

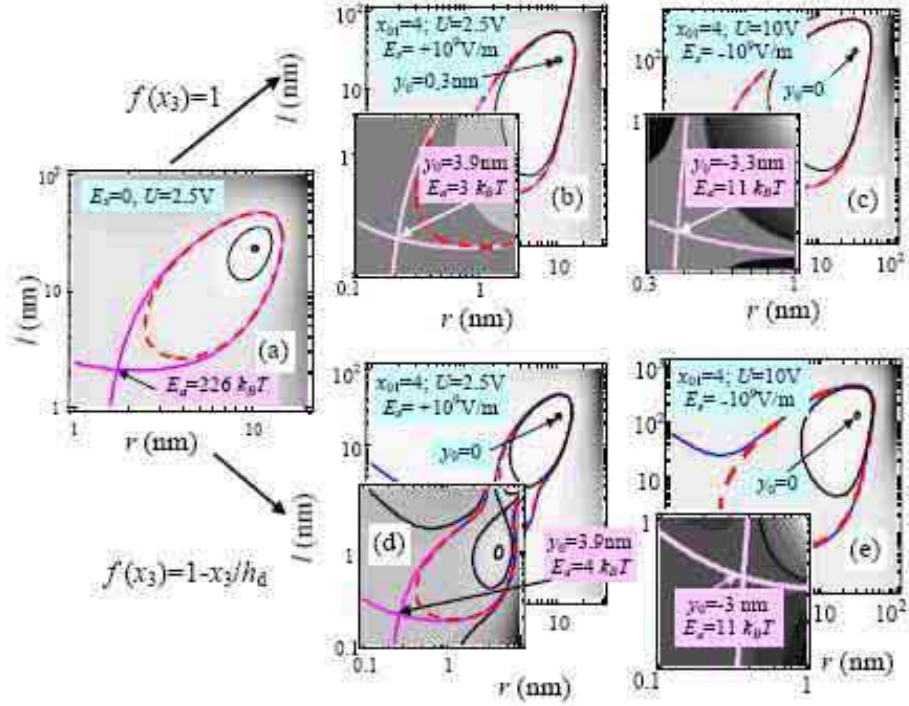

**FIG. 6**. Voltage dependence of free energy surfaces in the presence of surface field defect ($r_d = 4$ nm, $h_d = 0.8$ nm, $f = 1$ (top row) and $f = 1 - x_3/h_d$ (bottom row)) for different distances $x_{01}$ from the defect center and maximal field: $E_S = 0$ (a); $x_{01} = 4$ nm, attracting positive defect with $E_S = +10^9$ V/m (parts b, d) and $x_{01} = 4$ nm, repulsing negative defect with $E_S = -10^9$ V/m (parts c, e). Dashed contour corresponds to zero energy. Arrow with label and small circle denote activation barrier $E_a$ (saddle point and contour) and absolute minimum (stable domain) correspondingly. Material parameters and tip characteristics are the same as in Fig. 4.

Numerical simulations illustrate that the dislocation-type defects ($f = 1$) provide a more significant effect on domain nucleation and growth than dipole patch type defects ($f = 1 - x_3/h_d$). Hence, below we primarily consider only dislocation-type defects. The effect of domain nucleus attraction or repulsion by the surface field defect with center located at different distances $x_{01}$ from the probe apex is numerically analyzed in Fig. 7.

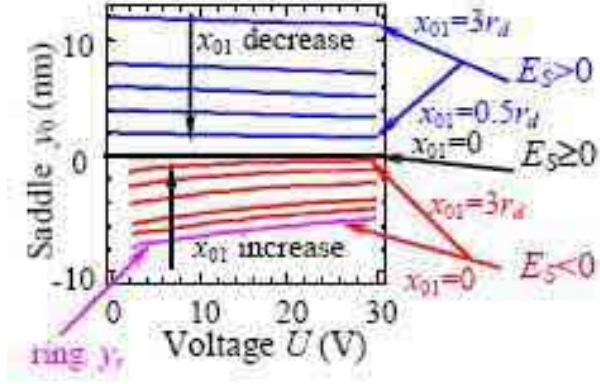

**FIG. 7**. (a) Voltage dependence of the domain nucleus center shift $y_0$ in the saddle point for surface field defect of $r_d = 4$ nm, $h_d = 0.8$ nm, $f = 1$ and field strength $E_S = +10^9; 0; -10^9$ V/m (see right labels). Shown are curves for tip-defect separations $x_{01} = 12; 8; 6; 4; 2; 0$ nm. Material parameters and tip characteristics are the same as in Fig. 4.

To describe the nucleus position analytically we performed minimization on $y_0$ of the free energy Eq. (15) at $f = 1$ under the conditions $r < 2d$, $l \leq 2\gamma d$ typically valid in a saddle point(s) and derived the set of approximate expressions for the shift $y_0$ (see Appendix B for more details):

$$y_0(U) \approx \begin{cases} \dfrac{x_{01} E_S \exp(-x_{01}^2/r_d^2)}{E_S \exp(-x_{01}^2/r_d^2) + U r_d^2(d+h_d)/2d^3 h_d}, & x_{01} \neq 0, \; UE_S > 0 \\[2ex] \dfrac{2\gamma d^3 h_d x_{01}}{r_d^2(d+h_d)} \dfrac{E_S}{U} \exp\left(-\left(2\gamma \dfrac{d^3}{r_d^2}\dfrac{E_S}{U}-1\right)^2 \dfrac{x_{01}^2}{r_d^2}\right), & x_{01} \neq 0, \; UE_S < 0 \\[2ex] r_d \sqrt{\ln\left(-\dfrac{2\sqrt{2}\gamma E_S d^3 h_d}{U r_d^2(d+h_d)}\right)}, & x_{01} = 0, \; 0 < U < -2\sqrt{2}\dfrac{\gamma E_S d^3 h_d}{r_d^2(d+h_d)} \\[2ex] 0, & x_{01} = 0, \; U > -2\sqrt{2}\dfrac{\gamma E_S d^3 h_d}{r_d^2(d+h_d)}. \end{cases}$$ (18)

Eqs. (18) qualitatively describe the behavior depicted in Fig. 7 including the cases of nucleus repulsion ($y_0 < 0$) at $E_S < 0$ and attraction ($y_0 > 0$) at $E_S > 0$ as well as $y_0 \to 0$ at $x_{01} \gg r_d$. As anticipated $y_0 \to 0$ at high voltages. For the stable domains with sizes $l \gg r$ and $l \gg h_d$ (typical for the tip-induced domain formation in the vicinity of a field defect with $|x_{01} - r_d| \lesssim d$) we obtained that $y_0 \approx 0$ for all voltages $U > \gamma |E_S| d^2 h_d / r_d^2$.

### IV.2. Activation barrier, critical voltage and domain sizes

For a favorable field defect ($E_S > 0$) domain nucleation can be either activationless at high enough built-in field or the activation barrier is lowered, rendering the process feasible at lower biases. For a negative field defect ($E_S < 0$) or its absence ($E_S = 0$) the domain formation process is always characterized by the activation energy, $E_a$, determined as the free energy value in the saddle point. Minimization of the free energy Eq. (15) on $r$ and $l$ under the conditions, $r < 2d$, $l < 2d$ typically valid at the nucleation stage (i.e. in a saddle point for $f = 1$) leads to the estimation of the activation barrier $E_a(U)$:

$$E_a(U) = \frac{2\pi\psi_S^3}{3}\left(\frac{3P_S d \cdot U}{\gamma\left(\sqrt{d^2+y_0^2}+d\right)^2} - \frac{P_S^2}{3\varepsilon_0\varepsilon_{11}} + \frac{3}{2}P_S E_S \exp\left(-\frac{(x_{01}-y_0)^2}{r_d^2}\right)F(h_d)\right)^{-2}. \quad (19)$$

Here the function $F(h_d) \cong \exp(-9\varepsilon_0\psi_S/8h_d P_S^2)$. Corresponding nucleus sizes are $l_S(U) \sim r_S(U) \cong \sqrt{3E_a(U)/2\pi\psi_S}$.

Following the definitions in Section III.1, activation voltages $U_a^0(E_S = 0)$ and $U_a^\pm(E_S \neq 0)$ corresponding to different polarization sign $\pm P_S$ (or, equivalently, forward and reverse switching) can be determined numerically from the free energy Eq. (15) using the conditions $\Phi(U_a^0, E_S = 0, \pm P_S, l_S, r_S) = E_a$ and $\Phi(U_a^\pm, E_S, \pm P_S, l_S, r_S) = E_a$ or estimated analytically from Eqs. (19). The following semi-quantitative approximations were derived for the defect-free case:

$$U_a^0 \cong \pm\frac{4}{3}\gamma d\left(\sqrt{\frac{2\pi\psi_S^3}{3E_a P_S^2}} + \frac{|P_S|}{3\varepsilon_0\varepsilon_{11}}\right) \quad (20a)$$

and defect-mediated switching:

$$U_a^\pm \cong \pm \frac{\gamma}{3d}\left(\sqrt{d^2+y_0^2}+d\right)^2\left(\sqrt{\frac{2\pi\psi_S^3}{3E_a P_S^2}}+\frac{|P_S|}{3\varepsilon_0\varepsilon_{11}}\right)-\Delta U(E_S) \qquad (20b)$$

$$\Delta U(E_S)=\frac{\gamma}{2d}\left(\sqrt{d^2+y_0^2}+d\right)^2 E_S \exp\left(-\frac{(x_{01}-y_0)^2}{r_d^2}\right)F(h_d) \qquad (20c)$$

Here $E_a$ is the potential barrier height chosen as a condition for thermally induced nucleation, e.g. 2-20$k_B$T. The lateral domain shift is

$$y_0 \cong \frac{x_{01}E_S \exp(-x_{01}^2/r_d^2)}{E_S \exp(-x_{01}^2/r_d^2)+U_a^0 r_d^2(d+h_d)/2d^3 h_d} \qquad (21a)$$

for $U_a^0 E_S > 0$, or

$$y_0 \cong \frac{2\gamma d^3 h_d x_{01}}{r_d^2(d+h_d)}\frac{E_S}{U_a^0}\exp\left(-\left(2\gamma\frac{d^3}{r_d^2}\frac{E_S}{U_a^0}-1\right)^2\frac{x_{01}^2}{r_d^2}\right) \qquad (21b)$$

for $U_a^0 E_S < 0$ (more rigorously, it could be estimated from Eq.(18) self-consistently).

From the analysis above, the effect of defect on the hysteresis loop shape can be predicted as follows. In the presence of a defect, the hysteresis loop is broadened by the factor $\left(\left(\sqrt{1+y_0^2/d^2}+1\right)^2-4\right)$ compared to the defect-free case. Furthermore, the loop is shifted along the voltage axis by the value $\Delta U$ due to domain-defect interactions. The value $\Delta U$ exponentially decreases with the distance $|x_{01}-y_0|$ from the defect center.

For a favorable field defect ($E_S > 0$) the domain nucleation with $P_S > 0$ is characterized by the smaller activation voltages, $U_a^+$, or can even be spontaneous (i.e. $U_a^+ = 0$, because $E_a(0) < 2k_B T$) at some values of $x_{01}$ and $E_S$. This corresponds to the case when the surface state already exists at zero voltage and a certain negative voltage $U_S^+$ is required to destroy it. Voltage dependence of the domain activation energy $E_a$ is shown in Fig. 8 (a). The estimation of the voltage $U_S^+$ can be obtained from the energy Eq. (19) as e.g. $E_a(U_S^+)=(2-20)k_B T$. Dependences of activation voltages $U_a^{0,\pm}$ (at levels 2 and 20$k_B$T) on the distance $x_{01}$ from the defect center are depicted in Fig. 8 (b).

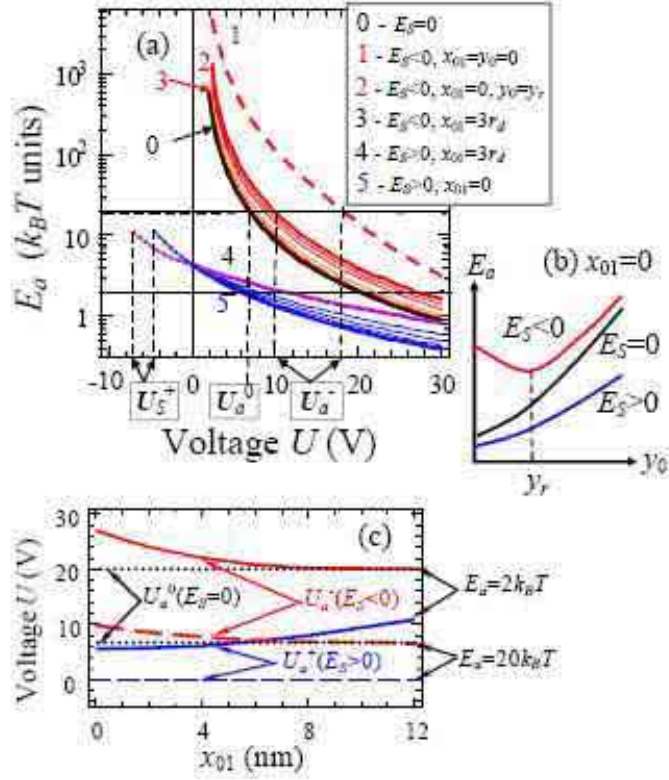

**FIG. 8**. (a) Dependence of the energy barrier (in $k_B T$ units) on the applied voltage $U$ for surface field defect of $r_d = 4$ nm, $h_d = 0.8$ nm, field strength $E_S$ located at position $x_{01}$. Curve 0 corresponds to $E_S = 0$. Curve 1 corresponds to the saddle $y_0 = 0$ (corresponding to the highest barrier, see schematics (b) for $E_a(y_0)$) at $E_S = -10^9$ V/m, $x_{01} = 0$; curve 2 corresponds to the saddle $y_0 = y_r \neq 0$ (corresponding to the lowest barrier, see schematics (c) for $E_a(y_0)$) at $E_S = -10^9$ V/m, $x_{01} = 0$; curve 3 is calculated at $E_S = -10^9$ V/m, $x_{01} = 3r_d$; curve 4 is calculated at $E_S = +10^9$ V/m, $x_{01} = 3r_d$; curve 5 is calculated at $E_S = +10^9$ V/m, $x_{01} = 0$. Intermediate thin curves are calculated at $x_{01} = 2r_d$; $1.5r_d$; $r_d$ and $0.5r_d$ nm correspondingly. (c) Dependence of the activation voltage $U_a$ at level 2 and 20$k_B$T on the distance to defect center, $x_{01}$. PZT-6B material parameters and tip characteristics are the same as in Fig. 4.

Similar analysis for the reversed domain nucleation with $P_S < 0$ affected by a negative surface field $E_S < 0$ requires the introduction of voltage $U_S^-$ corresponding to the surface state disappearance ($U_a^- = 0$ is possible).

For a material with PZT-6B parameters, the activation barrier may be extremely low in the vicinity, $x_{01}/r_d < 1$, of the positive surface field defect with sufficiently high field strength, $E_S > 10^8$ V/m. Curves 4-5 demonstrate that the surface state disappears at $U_S^+ \approx -5$ V. For a negative surface field defect ($x_{01}/r_d < 1$ and e.g. $E_S = -(10^8 - 10^{10})$ V/m) no surface state exists and the activation barrier drastically increases, as follows from curves 1-2. Thus, the surface field defect essentially facilitates or delays the tip-induced domain nucleation with respect to the activation voltage.

Comparing the data in Figs. 7-8 for $E_S > 0$ and $E_S < 0$, we conclude that the negative field defect influence is felt at larger distances, than the positive one at the chosen material parameters. However, the situation is general as it follows from Eqs. (18-19), since $|y_0(E_S > 0)| < |y_0(E_S < 0)|$ at $x_{01} \neq 0$ and $E_a(E_S > 0) < E_a(E_S < 0)$ always.

To further illustrate the defect-effect on local nucleation, we compare the influence of the defect field and location on the voltage dependence of equilibrium domain and nucleus sizes in Fig. 9.

From Fig. 9 (a), the equilibrium domain sizes are insensitive to the defect position and the field strength at the chosen material parameters are within the given range of defect sizes, $r_d$, $h_d$. Only the positions of the origins of the curves (corresponding to activation voltage $U_a^{-,0}$ or $U_S^+$) are sensitive to the defect characteristics. The reason for this behavior is the condition $U_a^- \gg U_{cr}^-$ ($U_a^- \gtrsim 20$ V and $U_{cr}^- \lesssim 3$ V). The critical voltage $U_{cr}$ depends on the defect characteristics, but it governs the thermodynamic domain formation only at a close activation barrier $U_a \sim U_{cr}$. At voltages $U \gg U_{cr}$ domain growth becomes almost independent of the initial critical point. In contrast, the bias dependence of nucleus sizes is sensitive to the surface field defect, as demonstrated in Fig. 9 (b,c). This analysis suggests that the primary influence of the surface field effect on the domain switching is the shift of activation energy (saddle point on free energy surface), while equilibrium domain size is almost unaffected.

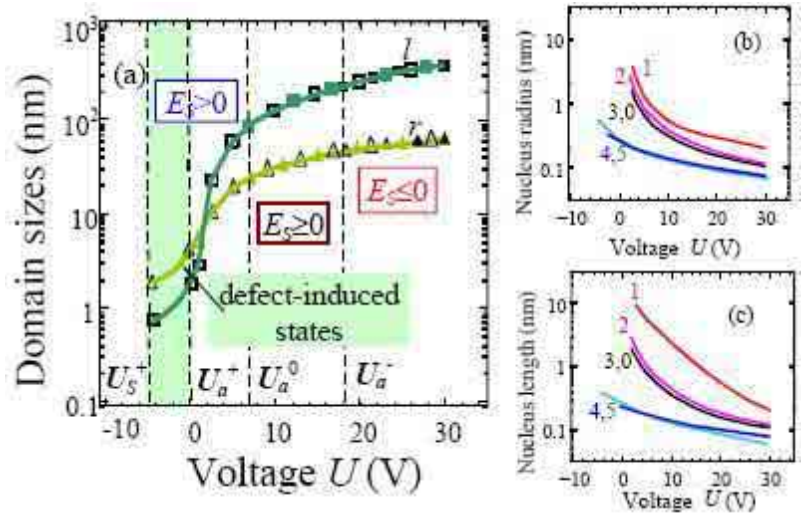

**FIG. 9**. (a) Voltage dependence of equilibrium domain radius $r$ and length $l$ on the applied voltage for different surface field $E_S$ : $E_S = +10^9$ V/m (empty symbols); $E_S = 0$ (color symbols) and $E_S = -10^9$ V/m (black symbols). (b,c) Voltage dependence of nucleus sizes in a saddle point. Curves 0-5 correspond to the same $E_S$ and $x_{01}$ values as described in Fig.8. PZT-6B material parameters, defect and tip characteristics are the same as in Figs. 4.

### IV.3. Effective piezoresponse and hysteresis loop fine structure

The effect of a surface field defect on the voltage dependence of the effective piezoresponse $d_{33}^{eff}(r(U))$ (i.e. local hysteresis loop) calculated from Eq. (10) is shown in Fig. 10 (a). Similarly to the behavior in Fig. 9 (a), only the starting points of the piezoresponse curves (voltages $U_a^0$, $U_a^\pm$ or $U_S^\pm$) are sensitive to the defect characteristics. However, this change in the nucleation voltage defines piezoresponse loop fine structure and horizontal asymmetry as shown in Figs. 10 (b-c).

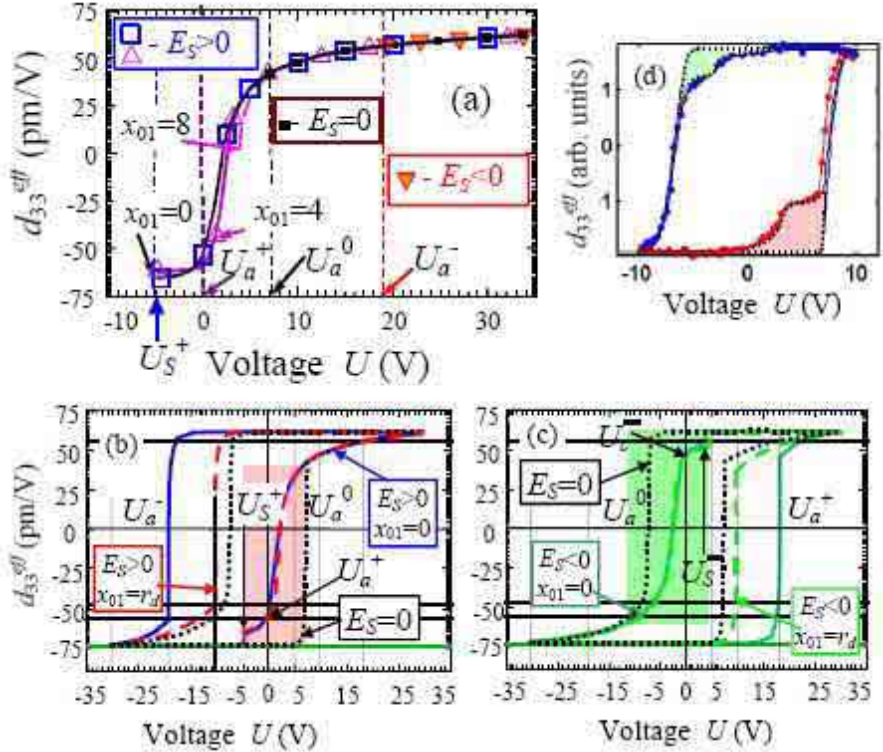

**FIG. 10**. Dependence of normalized PFM response (a) on the applied voltage affected by the surface field defect. (b, c) PFM response loops in the weak pinning limit at surface field amplitude $E_S = 0, +10^9$ V/m (b) and $E_S = 0, -10^9$ V/m (c). Dotted ($E_S = 0$), dashed ($x_{01} = r_d$, $E_S \neq 0$) and solid ($x_{01} = 0$, $E_S \neq 0$) curves are calculated for coercive biases $U_a^0$, $U_S^\pm$, $U_a^\pm$ estimated for nucleation onset $E_a = 20\,k_BT$. Piezoelectric coefficients $d_{15}$=135.6 pm/V, $d_{31}$=-28.7 pm/V, $d_{33}$=74.9 pm/V are used. (d) Typical experimental PFM response loop with fine structure (filled regions)[63]

For the above scenario, the positive (PNB) and negative (NNB) nucleation biases can be written as $U_a^+ = U_a^0 - \Delta U$ and $U_a^- = -U_a^0 - \Delta U$, correspondingly, where $U_a^0$ is the activation voltage that corresponds to defect-free nucleation (see symmetric dotted loops in Figs. 10b-c) and $\Delta U$ is described by Eq. (16b). The shift along the voltage axis is a direct effect of a defect influence. For $E_S > 0$, the nucleation bias can be zero, $U_a^+ = 0$, as shown in Fig. 10 (b). In this case, the piezoresponse loop exhibits fine structure at voltage $U_S^+$, i.e. a jump-like peculiarity corresponding to a delayed nucleation [the filled region in Fig. 10 (b)].

For a negative surface field, $E_S < 0$, the piezoresponse loop fine structure appeared at voltage $U_S^-$ and represents a bump-like peculiarity corresponding to rapid switching within the defect (the filled region in Fig. 10 (c)). Such loop fine structure is often observed on experimental data, as shown in Fig. 10 (d).

### V. Experimental observations of loop fine structure

The analysis performed in Sections III, IV suggests that the presence of the localized surface field defects can strongly affect the structure of the hysteresis loop in PFM, inducing significant asymmetry and introducing fine-structure features. Thus, the analysis of this fine structure and its variation from point to point on the surface can potentially provide information on the density and strength of the defect, i.e. allow the disorder potential to be reconstructed. While rigorous analysis will require numerical calculations due to the 3D nature of the problem, below we discuss the signatures of defects on PFM data and potential routes for semiquantitative data interpretation.

### V.1. Qualitative observation of complex structure in PFM hysteresis loops

The characteristic and easily identifiable signature of domain-defect interactions is the (reproducible) fine structure of hysteresis loops. The "non-ideal" loop shape can be noticed on many published examples of PFM spectroscopy, in some cases comparable or below the noise level. The work of Abplanalp[83] and Harnagea[84] attribute anomalous loop shapes to high-order switching and spatial confinement effects, respectively. An extensive number of anomalous loops were collected in work by Buhlmann.[85] The first report of anomalous loop shape as due to domain-ferroelastic wall interaction was published by Alexe et al. [62,86] and Jesse et al.[87]

In many cases, the recognition of the loop fine structure can be hindered by instrumental artifacts and noise. The advent of Switching Spectroscopy PFM allows arrays of hysteresis loops on a 2D surface mesh to be collected, thus allowing the reproducibility of the loop structure at a single point and systematic variations from pixel to pixel to be studied.[88] Simple examination of the spatial localization of hysteresis loop fine structure (e.g. Fig. 10 in Ref. [87]) illustrates that fine structure is correlated within a given region of the image, hence suggesting the presence of a local defect.

## V.2. Surface state maps and fine structure correlations

The analysis of SS-PFM 3D data sets allows the characteristic parameters describing polarization switching such as work of switching (area within the loop), positive and negative nucleation biases, positive and remnant coercive biases, etc. to be plotted as 2D maps, correlating switching behavior with local topography.

As discussed above, the characteristic feature of the field defect on the surface is the asymmetry in local nucleation bias. For rotationally invariant tips and well-separated defects (defect spacing >> defect size, tip size), the nucleation bias images are expected to exhibit well-defined circular features centered at the defect. The feature size is expected to be comparable to the defect size (intrinsic) or the tip radius (resolution limited). In the former case, the signal variation within the feature represents the internal structure of the defect, while in the latter the feature size is a measure of the probe size. This behavior is reported in Ref. [63]. Note that the difference in fine structures (one well-defined element for positive curve, several fine structure elements for negative) is consistent with the behavior in Fig. 8 (b), which illustrates that the effect of an attractive center is short ranged (defect attracts), while a repulsive defect is longer ranged. For dense defects (defect spacing is smaller than tip size), the individual signatures are not discernible, but the image will still illustrate correlations of the length scale of defect spacing.

To estimate the defect-mediated polarization switching for well-known materials, illustrated in Fig. 11-12 are dependences of activation voltage on the distance from the defect, defect maximum field, and defect radius in PZT-6B and BiFeO$_3$ (assuming effective tetragonal symmetry)[89].

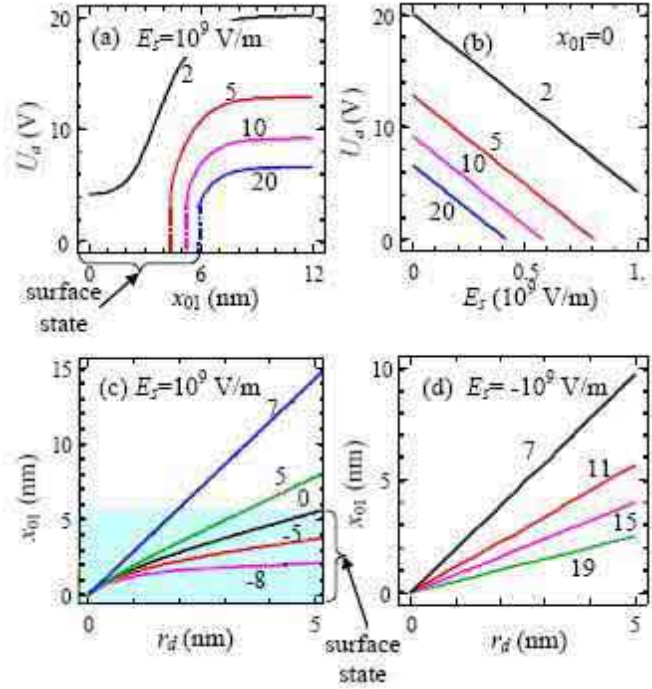

**FIG. 11**. Dependence of activation voltage on (a) the distance $x_{01}$ from the positive field defect (field strength $E_S = 10^9$ V/m, $r_d = 10$nm, $h_d = 0.8$ nm) and (b) maximum defect field strength $E_S$ for $x_{01} = 0$ in PZT-6B for different values of activation energy $E_a$ (figures near curves are $E_a$ values in $k_BT$ units). (c,d) Contour maps of $U_a$ via the distance $x_{01}$ and defect radius $r_d$ at $E_a = 20\, k_BT$ (figures near the curves are $U_a$ values in V) for field strength (c) $E_S = 10^9$ V/m and (d) $E_S = -10^9$ V/m. PZT-6B material parameters, tip characteristics are the same as in Fig. 4.

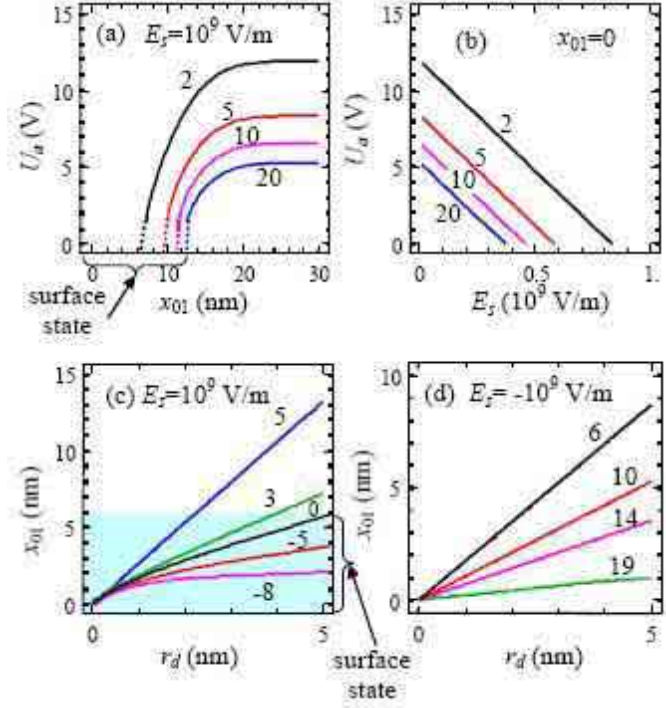

**FIG. 12**. Dependence of activation voltage $U_a$ on (a) the distance $x_{01}$ from the positive field defect (field strength $E_S = 10^9$ V/m $r_d = 10$nm, $h_d = 0.8$nm) and (b) maximum defect field strength $E_S$ for $x_{01} = 0$ in BiFeO$_3$ for different values of activation energy $E_a$ (figures near curves are $E_a$ values in $k_B T$ units). (c,d) Contour maps of $U_a$ via the distance $x_{01}$ and defect radius $r_d$ at $E_a = 20\,k_B T$ (figures near the curves are $U_a$ values in V) for field strength $E_S = 10^9$ V/m (c) and $E_S = -10^9$ V/m (d). Material parameters correspond to a tetragonal BiFeO$_3$: $P_S = 0.5$ C/m$^2$, $\varepsilon_{11} \approx \varepsilon_{33} \approx 80$, $\gamma \approx 1$, $\psi_S = 100$ mJ/m$^2$; point charge-surface separation $d = 7$ nm.

### V.3. Loop deconvolution and analysis of defects energetics

The semi-quantitative description of the piezoresponse loop $d_{33}^{eff}(U)$ fine structure requires several steps to be followed, including (i) tip shape calibration, (ii) deconvolution of the domain radius-voltage dependence $r(U)$, and (iii) analysis of defects energetics for a known set of $\{U_i, r_i\}$, where $U_i$ is a voltage corresponding to $i$-th fine feature, and $r_i$ corresponding domain size. For an ideal loop, $i = 1$.

Effective tip size (i.e. charge-surface distance $d$) can be determined self-consistently from the measured domain wall width as described in Ref [90]. With this information in hand, domain size deconvolution can be performed using an expression for $d_{33}^{eff}(r)$ given by Eqs. (10). However, even the approximate theoretical dependence $r(U)$ obtained after minimization of the free energy (12) is valid far from the critical point and rather cumbersome [see Eqs.(B.8-9) in Appendix B]. Hence, for a semi-quantitative analysis we propose the following procedure.

(a) Extracting the dependence $r(U)$ from the experimentally measured $d_{33}^{eff}(U)$ using Eqs. (10) at $y_0 = 0$, since we obtained that at voltages $U \geq U_a^+ > 0$ the domain center shift $y_0 \to 0$ even for the initial stages of domain formation and realistic surface field amplitudes $E_S \cong 10^9 - 10^{10}$ V/m.

(b) From the experimentally observed hysteresis loop $d_{33}^{eff}(U)$ asymmetry and fine structure at different tip location with respect to the defect position $x_{01}$ one extracts defect characteristics such as surface field amplitude, $E_S$, and defect radius, $r_d$. As a first approximation, well separated multiple field defects can be considered as a superposition of single ones (linear approximation).

Typical examples of BiFeO$_3$ hysteresis loops affected by growth defects are shown in Fig. 13. It is clear that $U_S^+$ values could be negative (a), approximately (b) zero or positive (c).

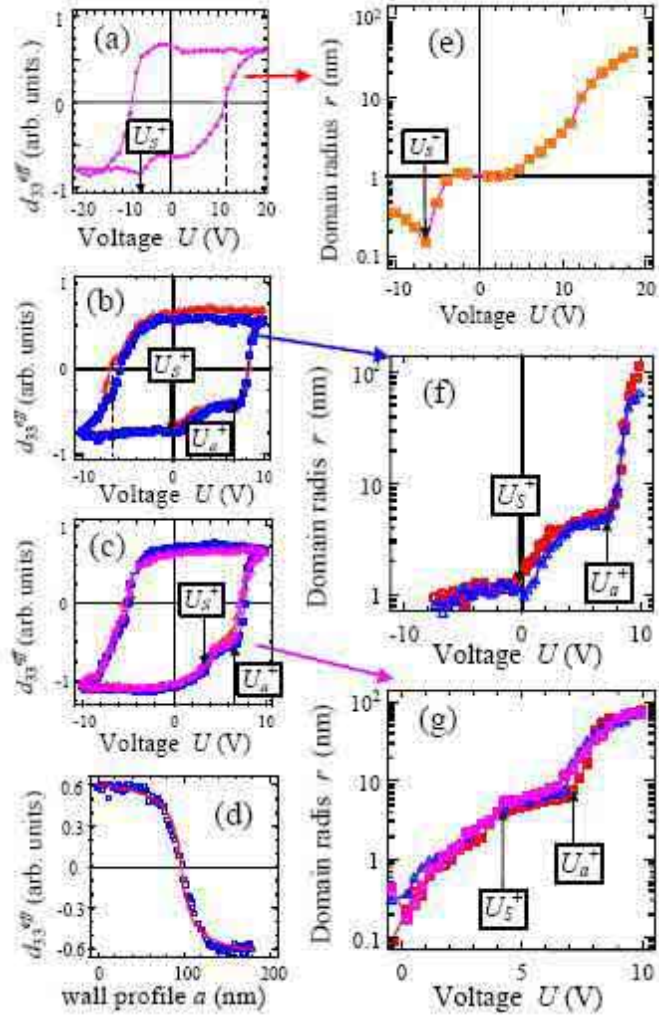

**FIG. 13**. (a, b, c) Normalized PFM response loops in 200nm-thick $BiFeO_3$ film with growth defects. (d) Tip calibration. Effective charge surface separation $d$=7 nm was calculated within the local point charge model $d = \varepsilon_e R_0/\kappa$ at $BiFeO_3$ permittivity $\kappa$ =80 and ambient permittivity $\varepsilon_e = 19$ obtained from the fitting of domain wall profile (d), where the fitted value $d$ =30 nm allows to obtain $\varepsilon_e = 19$ within the sphere-plane model $d = 2\varepsilon_e R_0 \ln((\varepsilon_e + \kappa)/2\varepsilon_e)/(\kappa - \varepsilon_e)$ at nominated tip curvature $R_0$=50 nm. (e, f, g) Voltage dependence of domain radius deconvoluted from the (a, b, c). For deconvolution the following parameters has been used: $d_{33}$=26 pm/V, $d_{15}$=3.5 pm/V, $d_{31}$=-12 pm/V, $\gamma = 1$.[91]

Deconvolution of the 3D data set of ferroelectric hysteresis loops acquired at each point of the image represents a complex problem, generally amenable only to numerical algorithms. Using Eqs. (20), we derive a simplified analytical model for the deconvolution of the nucleation bias and fine structure maps. For positive and negative nucleation bias maps for the case of nucleation below the tip, the nucleation biases are:

$$U_{an}^{\pm} \cong \pm U_a^0 - 2\gamma d \sum_i \tilde{E}_{di}\left(x_{1n} - x_{01}^i, x_{2n} - x_{02}^i\right) \tag{22}$$

Where $n = 1...N$ is the number of scanning points, $\{x_{1n}, x_{2n}\}$; $\{x_{01}^i, x_{02}^i\}$ is the center position of $i$-th surface field defect, $\tilde{E}_{di}(x_1, x_2)$; $\{x_{1n}, x_{2n}\}$ is the domain center position that coincides with the tip apex location.

From Eq.(22) the defect free nucleation bias

$$U_a^0 = \sum_n^N \frac{U_{an}^+ - U_{an}^-}{2N} \equiv \frac{4\gamma d}{3}\left(\sqrt{\frac{2\pi \psi_S^3}{3E_a P_S^2}} + \frac{|P_S|}{3\varepsilon_0 \varepsilon_{11}}\right) \tag{23}$$

and the bias difference (horizontal imprint bias)

$$\Delta U_{an} = U_{an}^+ + U_{an}^- = -4\gamma d \sum_i \tilde{E}_{di}\left(x_{1n} - x_{01}^i, x_{2n} - x_{02}^i\right). \tag{24}$$

For purely surface field defects, Eq. (24) allows the reconstruction of the electric fields of the defect directly from the SS-PFM imprint map, when separation $d$ is determined from the probe calibration and the relevant basis for the resulting field $\sum_i \tilde{E}_{di}(x_1, x_2)$ expansion is chosen. For the Gaussian basis considered the $i$-th defect surface field is $\tilde{E}_{di}(x_1, x_2) = \tilde{E}_{Si} \exp\left(-\frac{x_1^2 + x_2^2}{r_{di}^2}\right)$, where $\tilde{E}_{Si} = E_{Si} F(h_{di})$ is the field amplitude.

The voltages corresponding to the fine structure features are:

$$U_{Sn}^+ \cong \left(\sqrt{d^2 + y_{0n}^2} + d\right)^2 \left(U_a^0 - 2\gamma d \sum_i \tilde{E}_{di}\left(y_{01}^n - x_{01}^i, y_{02}^n - x_{02}^i\right)\right) \tag{25a}$$

$$U_{Sn}^- \cong -\left(\sqrt{d^2 + z_{0n}^2} + d\right)^2 \left(U_a^0 + 2\gamma d \sum_i \tilde{E}_{di}\left(z_{01}^n - x_{01}^i, z_{02}^n - x_{02}^i\right)\right) \tag{25b}$$

Where $\{y_{01}^n, y_{02}^n\}$ and $\{z_{01}^n, z_{02}^n\}$ are the domain center position that may differ from the tip apex location; $y_{0n}^2 = (y_{01}^n)^2 + (y_{01}^n)^2$ and $z_{0n}^2 = (z_{01}^n)^2 + (z_{01}^n)^2$.

In deconvolution of experimental data, nucleation and fine structure biases $\{U_{an}^-, U_{an}^+, U_{Sn}^-, U_{Sn}^+\}$ in the scanning points $\{x_{1n}, x_{2n}\}$ are determined from hysteresis loops and could be presented as local maps (see Figs. 14b-d).

- At the first step, the fitting is performed with respect to the set of parameters $\{x_{01}^i, x_{02}^i, \tilde{E}_{Si}, r_{di}\}$ determined using Gaussian fits from the imprint biases $\{\Delta U_{an}\}$ in the points $\{x_{1n}, x_{2n}\}$. The amount of defects (i.e. the number of basis elements) depends on the necessary accuracy of surface field reconstruction.
- At the second step, the domain center positions $\{y_{01}^n, y_{02}^n\}$ and $\{z_{01}^n, z_{02}^n\}$ are determined from the fine structure biases maps $\{U_{Sn}^-, U_{Sn}^+,\}$ using Gaussian fits.

The data in Fig. 14 (d) can be fitted using a model of 6 well-separated weak defects, or 3 strong defects, as demonstrated below.

Using experimental loops in a 200 nm BiFeO$_3$ film, partially shown in Fig. 13 (d), we have found that $U_a^0 \cong 5.5 \pm 0.3$ V and so $\psi_S = 104\,\text{mJ/m}^2$ in accordance with Eq. (23). The loop shapes are determined by three short-ranged negative defects: the first one "1" with field amplitude $\tilde{E}_{S1} = -800$ kV/cm, radius $r_{d1} = 6$nm located at coordinates {1.5nm, 7.5nm}; the second one "2" with $\tilde{E}_{S2} = -900$ kV/cm, $r_{d2} = 2$nm located at cell coordinates {15.5nm, 7.5nm}, the third one "3" with $\tilde{E}_{S3} = -700$ kV/cm, $r_{d3} = 3$nm located at cell coordinates {15nm, 0nm}.

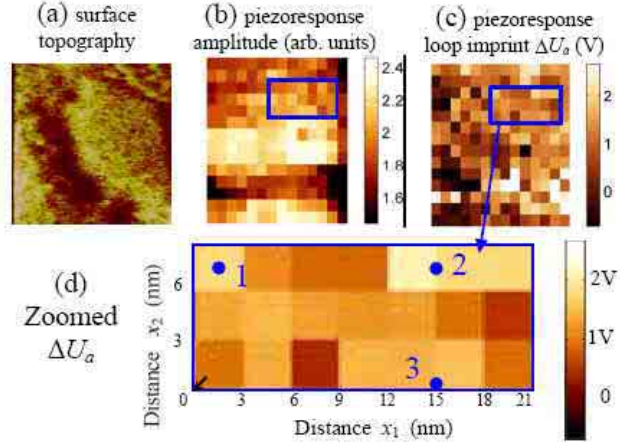

**FIG. 14**. (a) Surface topography of 200nm-thick BiFeO$_3$ film; (b) local map of surface piezoresponse amplitude; (c) local map of piezoresponse hysteresis loop imprint $\Delta U_a$ (d) Zoomed imprint $\Delta U_{an}$ with defects 1, 2 and 3 positions marked by circles (experimental scanning results in cell coordinates). Cell size is about 3 nm.

## V.4. Future prospects

The (semi)quantitative analysis developed above suggest that hysteresis loop fine structure at a single point and in the 3D SS-PFM arrays contains information on the defect-induced potential inside the material. While unambiguous analysis is possible only for a low density of defect sites and well-defined defect identities (e.g. surface field defect), the general form of Eqs. (20) suggests the possibility of the development of numerical schemes to extract the disorder potential in the general case. Ideally, this analysis will be based on the full loop shape. Furthermore, combination with synchrotron-based focused X-ray measurements will provide insight into the atomistic nature of the defects.

## VI. Conclusions

Here we have analyzed the effect of localized surface field defects on polarization switching in piezoresponse force microscopy using an extended Landauer-Molotskii model. The presence of the defect is shown to significantly affect the activation energy for loop formation. Depending on the relative sign of the defect field and tip potential, the defect can impede or facilitate the nucleation, resulting in significant asymmetry of the hysteresis loop. Remarkably, for the case studied here, the equilibrium domain size is not affected by the defect strength, and only the critical voltage required for switching is controlled by the defect, giving rise to universality of switching behavior.

The domain-defect interaction can result in the fine structure of the hysteresis loops, somewhat similar to the force-distance curves in force-based atomic force microscopy. Based on the thermodynamics of the switching process, the fine structure is expected to be reproducible at a single location, and vary on the length scale of defect size or tip size between adjacent spatial points. This behavior is found to be in agreement with broad array of experimental data on model $Pb(Zr,Ti)O_3$ and $BiFeO_3$ materials. The approaches for the deconvolution of the hysteresis loop fine structure and analysis of the defect parameters have been suggested.

The analysis presented here is performed for local polarization switching. However, it can similarly be extended to other voltage induced phase transitions, including amorphization-crystallization in phase change memories, bias-induced metal-insulator transitions, and electrochemical reactions. In these, bias induces local phase transformation, and locally measured signal provides the readout for the size of transformed region. The variation of tip location on the surface allows the switching behavior to be correlated with microstructure. Giving the role bias –induced phase transitions play in information technology (operation of virtually all electronic devices is based on the interaction between electrical bias and matter) and energy-related research, the capability to probe the role of local defects on these phenomena is a key to future progress. For electromechanically active materials such as ferroelectrics and multiferroics, polyelectrolytes, biopolymers, redox active molecules, and biological systems, the detection method can be based on local electromechanical response. For other systems, tip-surface current or tip-assisted Raman and near-field optical measurements provide an additional channel of information. The comparison of the force-based and bias-based methods is given in Table I.

Finally, the primary limitation of functional SPM imaging is a lack of information on the atomic identity of the local defects. The combination of SPM with in-situ electron microscopy or the use of systems with engineered domain structures (e.g. bicrystal grain boundaries or periodic dislocation network arrays) offers model systems with well-defined defect sites. These combinations will allow correlation of defect mediated thermodynamics and kinetics of phase transitions with atomic structure, paving the pathway for understanding the atomistic mechanisms of switching.

**Table I.** Comparison of force-induced and bias-induced phase transitions

|  | Protein unfolding | Nanoindentation | Bias-induced |
|---|---|---|---|
| Energy scale | 0.1 eV | $>10^4$ eV | ~1-100 eV |
| Reversibility | reversible in certain cases | irreversible in plastic regime | Reversible for ferroelectric switching, generally irreversible if includes mass, exchange |
| Applicability | Proteins, DNA, etc. | All materials | Ferroelectrics, piezoelectric inorganic and biomaterials, redox-active systems |
| Notes | Require molecule hunting | One location only | 2 disorder potential components |

## Appendix A. Field defect charge density determination

We define the coordinate system with the origin at the defect center $\{x_{01},0,0\}$ and use the cylindrical coordinate system $\{\rho,\varphi,z\}$. The bulk charge density $\sigma_d(\rho,z)$ induced by the axisymmetric field defect with a given z-component of the electric field, $E_z^S(\rho,z)$, can be found from the system of electrostatic Maxwell equations $rot\,\mathbf{E}^S = 0$, $\varepsilon_0 div(\hat{\varepsilon}\mathbf{E}^S) = \sigma_d$. In cylindrical coordinates $\{\rho,\varphi,z\}$ we obtain:

$$\frac{\partial}{\partial z}E_\rho^S(\rho,z) = \frac{\partial}{\partial \rho}E_z^S(\rho,z), \quad E_\rho^S(\rho,z) = \int^z \frac{\partial}{\partial \rho}E_z^S(\rho,z')dz', \tag{A.1}$$

$$\sigma_d(\rho,z) = \varepsilon_0\left(\frac{\varepsilon_{11}}{\rho}\frac{\partial}{\partial \rho}\left(\rho E_\rho^S(\rho,z)\right) + \varepsilon_{33}\frac{\partial}{\partial z}E_z^S(\rho,z)\right). \tag{A.2}$$

Eq. (A.1) should be supplemented with relevant boundary conditions. For the perfect electric contact between the conductive tip and surface $E_\rho^S(\rho,z=0) = 0$ and $-\varepsilon_{33}\varepsilon_0 E_3^S(\rho,z=0) = \sigma_b(\rho)$. The same model was used for depolarization and interaction energy calculations [e.g. Eq.(15)]. The free charge density $\sigma_b(\rho)$ is located inside the screening layer or flattened tip apex or top electrode. Further derivation depends on the expression for $E_z^S(\rho,z)$.

*Case (a):* z-component of the electric field is

$$E_3^S(x,y,z) = E_S \exp\left(-\frac{(x-x_{01})^2 + y^2}{r_d^2} - \frac{z}{h_d}\right). \tag{A.3}$$

Corresponding defect charge density is

$$\sigma_d(x,y,z) = -\varepsilon_0 E_3^S(x,y,z)\frac{\varepsilon_{33}r_d^4 + 4\varepsilon_{11}(\exp(z/h_d)-1)h_d^2\left(r_d^2 - (x-x_{01})^2 - y^2\right)}{h_d r_d^4} \tag{A.4}$$

The charge density is maximal near the surface and rapidly decreases with depth, z. However it tends to constant value at $z \to \infty$, namely

$$\sigma_S(x,y,\infty) = -4\varepsilon_0\varepsilon_{11}E_S \exp\left(-\frac{(x-x_{01})^2 + y^2}{r_d^2}\right)\left(r_d^2 - (x-x_{01})^2 - y^2\right)\frac{h_d}{r_d^4} \tag{A.5}$$

Thus Eq.(A.4) describes a continuous distribution resembling that for a charged dislocation line, in agreement with analysis by Weber et al.[81]

*Case (b):* $z$-component of the electric field is

$$E_3^S(x, y, z) = E_S \exp\left(-\frac{\sqrt{(x - x_{01})^2 + y^2}}{r_d} - \frac{z}{h_d}\right). \tag{A.6}$$

Under the condition $E_\rho^S(\rho, z = 0) = 0$ (corresponding to adopted model) the defect charge density is

$$\sigma_S(x, y, z) = -\varepsilon_0 E_3^S(x, y, z) \left(\frac{\varepsilon_{33}}{h_d} + \left(\frac{\varepsilon_{11} h_d}{r_d^2} - \frac{\varepsilon_{11} h_d}{r_d \sqrt{(x - x_{01})^2 + y^2}}\right)\left(\exp\left(\frac{z}{h_d}\right) - 1\right)\right) \tag{A.7}$$

The distribution (A.7) is maximal near the surface and rapidly decreases with depth $z$ decrease, except singularity along the vertical charged line $\{x_{01}, 0, z\}$. Furthermore, expression (A.7) tends to the constant value at $z \to \infty$, namely

$$\sigma_S(x, y, \infty) = -\varepsilon_0 \varepsilon_{11} E_S \exp\left(-\frac{(x - x_{01})^2 + y^2}{r_d^2}\right)\left(\frac{h_d}{r_d^2} - \frac{h_d}{r_d \sqrt{(x - x_{01})^2 + y^2}}\right), \tag{A.8}$$

Charge distributions calculated from Eqs.(A.7) are much weaker localized in the transverse direction in comparison with the ones given by Eqs.(A.4), as anticipated.

*Case (c):* The defect potential

$$\varphi^S(x, y, z) = -E_S z \exp\left(-\frac{(x - x_{01})^2 + y^2}{r_d^2} - \frac{z}{h_d}\right) \tag{A.9}$$

that satisfies the boundary condition $\varphi^S(x, y, 0) = 0$ for the perfect tip-surface electric contact. Corresponding built-in field and defect charge density are:

$$E_3^S(x, y, z) = E_S\left(1 - \frac{z}{h_d}\right)\exp\left(-\frac{(x - x_{01})^2 + y^2}{r_d^2} - \frac{z}{h_d}\right) \tag{A.10}$$

$$\sigma_d(x, y, z) = \varepsilon_0 E_3^S(x, y, z) \frac{\varepsilon_{33} r_d^4 (z - 2h_d) + 4\varepsilon_{11} z h_d^2 \left((x - x_{01})^2 + y^2 - r_d^2\right)}{h_d^2 r_d^4} \tag{A.11}$$

Expression (A.11) is maximal near the surface and exponentially vanishing with depth $z$ decrease. It is well-localized charge spot that produces electric field with different polarity.

## Appendix B. Approximate analytical analyses

*B.1. Saddle point*

To obtain approximate analytical results we simplify the free energy Eq. (15) under the conditions, $r < 2d$, $l < 2\gamma d$ typically valid at nucleation stage (i.e. in a saddle point).

For $f = 1$ and $l < h_d$ and $r < r_d$ the free energy (15)-(16) Pade approximation becomes:

$$\Phi \approx f_S r^2 S_D(\xi) + f_D r^3 N_D(\xi) + (f_U + f_d) r^3 / \xi \qquad (B.1a)$$

Note that Eq.(B.1a) is valid at $d < h_d$ (since $r_d \gg h_d$).

For $r \gg r_d$ and $l \gg h_d$ it acquires the form

$$\Phi \approx f_S r^2 S_D(\xi) + f_D r^3 N_D(\xi) + f_U r^3 / \xi + f_d h_d r_d^2 \qquad (B.1b)$$

For $f = 1 - x_3/h_d$ the free energy (15)-(16) Pade approximation becomes:

$$\Phi \approx f_S r^2 S_D(\xi) + f_D r^3 N_D(\xi) + \left( f_U + f_d \exp\left(-\frac{r}{\xi h_d}\right) \right) r^3 / \xi \qquad (B.1c)$$

In Eqs (B.1) the domain aspect ratio $\xi = r/l$ determines the shape-function

$$S_D(\xi) = 1 + \frac{\arcsin\sqrt{1-\xi^2}}{\xi\sqrt{1-\xi^2}}$$

and depolarization factor

$$N_D(\xi) = \frac{\xi \gamma^2}{1-\xi^2\gamma^2}\left( \frac{\operatorname{arctanh}\left(\sqrt{1-\xi^2\gamma^2}\right)}{\sqrt{1-\xi^2\gamma^2}} - 1 \right)$$

varying within the range $\{0,1\}$ in SI. The characteristic energies are $f_S = \pi\psi_S$, $f_D = \dfrac{4\pi P_S^2}{3\varepsilon_0\varepsilon_{11}}$, $f_U(U) = -\dfrac{4\pi P_S d\, U}{\gamma\left(\sqrt{d^2 + y_0^2} + d\right)^2}$,

$f_d = -2\pi P_S E_S \exp\left(-\dfrac{(x_{01} - y_0)^2}{r_d^2}\right)$. The nucleus center shift $y_0$ should satisfy transcendental equation $\dfrac{d(f_U + f_d)}{dy_0} = 0$.

For $f = 1$ minimization of Eq.(B.1a) on the variables $r$ and $\xi$ leads to the parametric dependences on $\xi$ of domain radius $r$, length $l$ and activation energy $E_a$:

$$r(U) = \frac{-2f_s \xi S_D(\xi)}{3(f_U + f_d + f_D \xi N_D(\xi))}, \tag{B.2a}$$

$$l(U) = \frac{-2f_s S_D(\xi)}{3(f_U + f_d + f_D \xi N_D(\xi))}, \tag{B.2b}$$

$$E_a(U) = \frac{-4f_s^3 \xi^2 S_D(\xi)^3}{27(f_U + f_d + f_D \xi N_D(\xi))^2}. \tag{B.2c}$$

Transcendental equation for the parameter $\xi$ has the form:

$$\xi^2 \frac{2S_D(\xi)(dN_D(\xi)/d\xi) - 3N_D(\xi)(dS_D(\xi)/d\xi)}{2S_D(\xi) + 3\xi(dS_D(\xi)/d\xi)} = \frac{f_U + f_d}{f_D} \tag{B.3}$$

At $\xi \to 1$ denominator $2S_D(\xi) + 3\xi(dS_D(\xi)/d\xi)$ tends to zero, so that for the case of high biases $-(f_U + f_d) \gg f_D$ one can obtain approximate expressions for $\xi$. For the corresponding $S_D(\xi)$ and $N_D(\xi)$ the asymptotic representation is $\xi \approx 1 + \frac{f_D}{2(f_U + f_d)}$ (see Fig.1B).

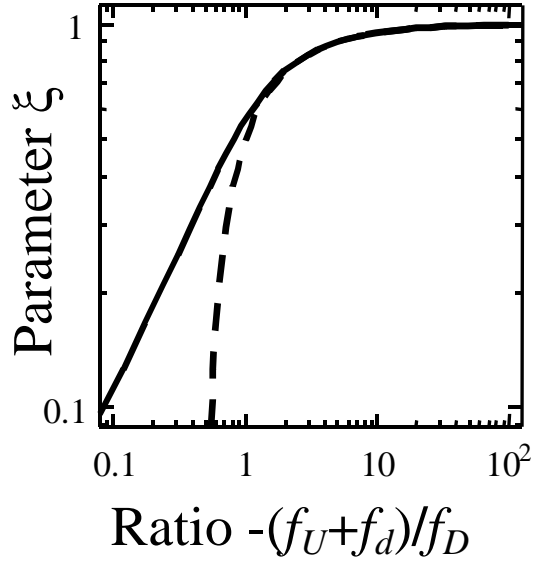

Fig. 1B. Parameter $\xi$ via the ratio $-(f_U + f_d)/f_D$. Solid curve is exact solution, dashed curve is approximation $\xi \approx 1 + \frac{f_D}{2(f_U + f_d)}$.

The expression along with Eqs.(B.2) leads to

$$r(U) \approx \frac{-4 f_S}{3(f_U + f_d) + f_D},$$  (B.4a)

$$l(U) \approx \frac{-4 f_S}{3(f_U + f_d) + f_D}\left(1 - \frac{f_D}{2(f_U + f_d)}\right),$$  (B.4b)

$$E_a(U) \approx \frac{32(f_S)^3}{3(3(f_U + f_d) + f_D)^2}.$$  (B.4c)

The Eqs. (B.2-4) are valid for $l \leq h_d$ and $r \leq r_d$ (by definition $r_d \gg h_d$).

Consideration of the free energy (B.1b) valid in the opposite case $l \gg h_d$ and $r \gg r_d$ leads to the same functional dependencies (B.2-4), where $f_U + f_d \to f_U$. Taking into account that the overlap integral (16) derivatives on $l$ exponentially vanishes as $\exp(-l/h_d)$ with domain length increase and the critical nucleus length estimated under the condition $f_U + f_d > -f_D$ is about $1.5(f_S/f_D)$, expressions particular cases $l \leq h_d$ and $l \gg h_d$ can be joined by substitution $f_d \to f_d \exp\left(-\frac{3 f_S}{2 f_D h_d}\right)$. Thus, substituting the characteristic energies in Eqs.(B.4) yields:

$$r_S(U) \approx \psi_S \left(\frac{3 P_S d \cdot U}{\gamma\left(\sqrt{d^2 + y_0^2} + d\right)^2} - \frac{P_S^2}{3\varepsilon_0 \varepsilon_{11}} + \frac{3}{2} P_S E_S \exp\left(-\frac{(x_{01} - y_0)^2}{r_d^2}\right) F(h_d)\right)^{-1},$$  (B.5a)

$$l_S(U) \approx r_S \left(1 + \frac{P_S}{3\varepsilon_0 \varepsilon_{11}}\left(\frac{2 d U}{\gamma\left(\sqrt{d^2 + y_0^2} + d\right)^2} + E_S \exp\left(-\frac{(x_{01} - y_0)^2}{r_d^2}\right) F(h_d)\right)^{-1}\right),$$  (B.5b)

$$E_a(U) \approx \frac{2\pi \psi_S^3}{3}\left(\frac{3 P_S d \cdot U}{\gamma\left(\sqrt{d^2 + y_0^2} + d\right)^2} - \frac{P_S^2}{3\varepsilon_0 \varepsilon_{11}} + \frac{3}{2} P_S E_S \exp\left(-\frac{(x_{01} - y_0)^2}{r_d^2}\right) F(h_d)\right)^{-2}.$$  (B.5c)

Here the function $F(h_d) = \exp\left(-\frac{9\varepsilon_0 \psi_S}{8 h_d P_S^2}\right)$, where the thickness $\varepsilon_0 \psi_S / P_S^2$ is proportional to the intrinsic domain wall width, of the order of several lattice constants. It reflects the fact that the critical domain sizes cannot be smaller than the width.

*B.2. Domain center shift*

Minimization of the free energy (B.1a) on $y_0$ leads to the transcendental equations:

$$y_0(U) = \frac{x_{01} E_S \exp\left(-(x_{01} - y_0)^2 / r_d^2\right)}{E_S \exp\left(-(x_{01} - y_0)^2 / r_d^2\right) + 4U r_d^2 d / \gamma \left(\sqrt{d^2 + y_0^2} + d\right)^3 \sqrt{d^2 + y_0^2}} \quad (B.6a)$$

Eq. B(6a) can be rewritten as:

$$\frac{y_0 - x_{01}}{r_d} \exp\left(-\left(\frac{y_0 - x_{01}}{r_d}\right)^2\right) = \frac{-4U r_d y_0 d}{E_S \gamma \left(\sqrt{d^2 + y_0^2} + d\right)^3 \sqrt{d^2 + y_0^2}} \quad (B.6b)$$

In special case $x_{01} = 0$, Eqs.(B.6) give two possibilities: $y_0 = 0$ for $E_S U > 0$ and the special point $y_0 = y_r \neq 0$ corresponding to the divergence of denominator and existing at $E_S U < 0$, namely:

$$y_r = \pm r_d \sqrt{\ln\left(-\frac{\gamma E_S \sqrt{d^2 + y_r^2}}{4U r_d^2 d}\left(\sqrt{d^2 + y_r^2} + d\right)^3\right)} \approx \pm r_d \sqrt{\ln\left(-2\sqrt{2}\gamma \frac{E_S d^3}{U r_d^2}\right)} \quad (B.7)$$

When the denominator in Eq.(B.6a) is finite, it includes the cases $y_0 < 0$ at $UE_S < 0$ and $y_0 > 0$ at $UE_S > 0$. So, under the condition $x_{01} \neq 0$ approximate expression for the shift $y_0(U, r_s, l_s)$ is:

$$y_0 \approx \begin{cases} \dfrac{x_{01} E_S \exp\left(-x_{01}^2 / r_d^2\right)}{E_S \exp\left(-x_{01}^2 / r_d^2\right) + U r_d^2 / 2\gamma d^3}, & E_S U > 0 \\ 2\gamma \dfrac{d^3}{r_d^2} \dfrac{x_{01}}{U} E_S \exp\left(-\left(2\gamma \dfrac{d^3}{r_d^2} \dfrac{E_S}{U} - 1\right)^2 \dfrac{x_{01}^2}{r_d^2}\right), & E_S U < 0 \end{cases} \quad (B.8)$$

Note that Eqs.(B.7)-(B.8) are derived for the case $l < h_d$ and $l < \gamma d$. Also we consider the case $l \gg h_d$, $l \gg \gamma d$ and obtained after elementary transformations:

$$y_r \approx \pm r_d \sqrt{\ln\left(-2\sqrt{2}\gamma \frac{E_S d^2 h_d}{U r_d^2}\right)} \quad (B.9)$$

Under the condition $x_{01} \neq 0$ and $E_S > 0$:

$$y_0 \approx \frac{x_{01} E_S \exp(-x_{01}^2/r_d^2)}{E_S \exp(-x_{01}^2/r_d^2) + U r_d^2 / 2d^2 h_d}$$ (B.10)

Expressions (B.8) and (B.10) can be joined together in the sense of Pade approximation, as proposed in the main text (see Eqs.(18)).

The kinetic instability corresponding to the switching between these two saddles ($y_0 = 0$ and $y_0 = y_r$) is possible, while in thermodynamic limit the one corresponding to the lowest activation energy is realized.

### B.3. Surface state critical field

From symmetry considerations, $x_{01} = y_0$ under the absence of external voltage $U$. For the case, numerical simulations proved that spontaneous (i.e. activationless) domain appearance is possible at $P_S E_S > 0$. At $U = 0$ the free energy is

$$f_S S_D(\xi) r^2 + f_D N_D(\xi) r^3 + f_d h_d r_d^2 \left(1 - \exp\left(-\frac{r^2}{r_d^2}\right)\right)\left(1 - \exp\left(-\frac{r}{\xi h_d}\right)\right)$$ for the case $f = 1$ and

$$f_S S_D(\xi) r^2 + f_D N_D(\xi) r^3 + f_d r_d^2 \frac{r}{\xi}\left(1 - \exp\left(-\frac{r^2}{r_d^2}\right)\right) \exp\left(-\frac{r}{\xi h_d}\right)$$ for $f = 1 - x_3/h_d$. Typically,

the spontaneous domain appears with sizes $l \leq h_d$, $r \leq r_d$ and $l < r$ (since $h_d \ll r_d$) if the built-in field more than the critical value: $E_S^{cr} \approx \frac{2}{3}\left(\frac{\psi_S}{h_d P_S} + \frac{P_S}{\varepsilon_0 \varepsilon_{11}}\right)$. More rigorous estimation of the critical built-in field of surface state appearance leads to

$$E_S^{cr} \cong \begin{cases} \dfrac{2e}{3(e-1)}\left(\dfrac{\psi_S}{h_d P_S} + \dfrac{P_S}{\varepsilon_0 \varepsilon_{11}}\right), & f = 1 \\ \\ \dfrac{2}{3e}\left(\dfrac{\psi_S}{h_d P_S} + \dfrac{P_S}{\varepsilon_0 \varepsilon_{11}}\right), & f = 1 - \dfrac{x_3}{h_d} \end{cases}.$$ (B.11)

The voltage of the saddle point appearance (preceding to the critical point or surface state origin) could be estimated from Eq.(B.5c) as $E_a \leq 2 - 20 k_B T$ allowing for the condition $y_0 \sim x_{01}$ at $U_S^{\pm} E_S < 0$. Note, that the voltage could be negative indicating the possibility of surface state (meta)stability.

Under the condition $E_S > 0$, the jump appeared at voltage $U_S^+ < 0$ ($U_a^+ = 0$) of surface domain state origin

$$U_S^+ \cong \frac{\gamma}{3d}\left(\sqrt{d^2+y_0^2}+d\right)^2 \left(\sqrt{\frac{2\pi\psi_S^3}{3E_a P_S^2}}+\frac{|P_S|}{3\varepsilon_0\varepsilon_{11}}-\frac{3}{2}E_S \exp\left(-\frac{(y_0-x_{01})^2}{r_d^2}\right)F(h_d)\right) \quad (B.12a)$$

Under the condition $E_S < 0$, the jump appeared at voltage $U_S^- > 0$ ($U_a^- = 0$) of surface domain state appearance

$$U_S^- \cong -\frac{\gamma}{3d}\left(\sqrt{d^2+y_0^2}+d\right)^2 \left(\sqrt{\frac{2\pi\psi_S^3}{3E_a P_S^2}}+\frac{|P_S|}{3\varepsilon_0\varepsilon_{11}}+\frac{3}{2}E_S \exp\left(-\frac{(y_0-x_{01})^2}{r_d^2}\right)F(h_d)\right) \quad (B.12b)$$

Here $E_a$ is the activation energy chosen as condition for thermally induced nucleation, e.g. 2-20$k_B$T.